\documentclass[aps,prb,reprint,superscriptaddress]{revtex4-2}
\usepackage[dvipsnames]{xcolor}
\usepackage{graphicx}
\usepackage{amsmath,amssymb}
\usepackage{bm}
\usepackage{braket}
\usepackage{lipsum}
\usepackage{makerobust}
\usepackage{simpler-wick}
\usepackage{hyperref}
\usepackage{comment}
\usepackage{bbold}

\newcommand{\makeauthor}[2]{\newcommand{#1}[1]{{%
  \sffamily\color{#2}{%
    \bfseries\begingroup\escapechar=-1\edef\x{\endgroup\string#1}\x:%
  } ##1}}%
  \MakeRobustCommand#1}
\makeauthor{\cole}{Plum}
\makeauthor{\themba}{ForestGreen} 
\makeauthor{\sr}{blue}

\begin{document}

\renewcommand{\vec}[1]{\bm{#1}}
\newcommand{\up}{{\uparrow}}
\newcommand{\dw}{{\downarrow}}
\newcommand{\pa}{{\partial}}
\newcommand{\pd}{{\phantom{\dagger}}}
\newcommand{\bs}[1]{\boldsymbol{#1}}
\newcommand{\add}[1]{{{\color{blue}#1}}}
\newcommand{\todo}[1]{{\textbf{\color{red}ToDo: #1}}}
\newcommand{\tbr}[1]{{\textbf{\color{red}\underline{ToBeRemoved:} #1}}}
\newcommand{\eps}{{\varepsilon}}
\newcommand{\nn}{\nonumber}
\newcommand{\Pf}{\text{Pf}}
\def\ie{\emph{i.e.},\ }
\def\eg{\emph{e.g.},\ }
\def\ea{\emph{et. al.}\ }
\def\cf{\emph{cf.}\ }

\newcommand{\brap}[1]{{\bra{#1}_{\rm phys}}}
\newcommand{\bral}[1]{{\bra{#1}_{\rm log}}}
\newcommand{\ketp}[1]{{\ket{#1}_{\rm phys}}}
\newcommand{\ketl}[1]{{\ket{#1}_{\rm log}}}
\newcommand{\braketp}[1]{{\braket{#1}_{\rm phys}}}
\newcommand{\braketl}[1]{{\braket{#1}_{\rm log}}}
\graphicspath{{./}{./figures/}}

\title{Negative Hybridization:\\
a Potential Cure for Braiding with Imperfect Majorana Modes}

\author{Cole Peeters}
\affiliation{School of Physics, University of Melbourne, Parkville, VIC 3010, Australia}
\author{Themba Hodge}
\affiliation{School of Physics, University of Melbourne, Parkville, VIC 3010, Australia}
\author{Stephan Rachel}
\affiliation{School of Physics, University of Melbourne, Parkville, VIC 3010, Australia}

\begin{abstract}
Majorana zero modes, the elementary building blocks for the quantum bits of topological quantum computers, are known to suffer from hybridization as their wavefunctions begin to overlap.
This breaks the ground state degeneracy, splitting their energy levels and leading to an accumulation of error when performing topological quantum gates. 
Here we show that the energy splitting of the Majorana zero modes can become {\it negative}, which can be utilized to reduce the average hybridization energy of the total gate. We present two illustrative examples where {\it negative hybridization} suppresses gate errors to such an extent that they remain below the fault-tolerance threshold. As an intrinsic property of Majorana zero modes, negative hybridization enables systems based on imperfect Majorana zero modes to regain functionality for quantum information processing.
\end{abstract}

\maketitle
%
%
\textit{Introduction.}---Topological quantum computing promises fault-tolerance against local perturbations, as quantum gates are enacted by pairwise exchanges of spatially separated Majorana modes, referred to as braids\,\cite{Moore1991,Ivanov2001,nayak2008,Elliott2015,beenakker20,jackDetectingDistinguishingMajorana2021,Yazdani2023}. 
A topological qubit consists of four Majorana zero modes (MZMs), and the non-locality of such a qubit is key for its topological protection\,\cite{nayak-96npb529,dassarma-15npjqi15001}. 
A major research question revolves around the actual fault tolerance of a Majorana-based quantum computer\,\cite{freedman-03bams31}. 
It is well established that the successful execution of a quantum gate 
necessitates that the corresponding braiding time, $T$, satisfies the ``speed limit'' of Majorana modes\,\cite{nayak2008,Brouwer2011,hodgeCharacterizingDynamicHybridization2025}:
\begin{equation}\label{eq:speedlimit}
    \frac{\hbar}{\Delta_{\rm topo}} \ll T \ll \frac{\hbar}{E_{\rm hyb}}\ ,
\end{equation}
where $\Delta_{\rm topo}$ is the size of the topological gap and $E_{\rm hyb}$ is the hybridization energy of a pair of Majorana modes.

The lower speed limit, 
$\hbar/\Delta_{\rm topo}$, protects the Majorana subspace from transitions into, or coupling to, bulk states. 
Intuitively, if the Majorana modes are moved too fast, they may undergo transitions to higher (excited) states, thus losing quantum information. 
The smaller the gap, the greater the likelihood of this diabatic error occurring\,\cite{karzig2013,Scheurer2013,Knapp2016}. 
The obvious way to escape from such errors is to perform the dynamics of the Majorana modes sufficiently slowly, i.e., adiabatically.

The source of the upper speed limit, $\hbar/E_{\rm hyb}$, is less obvious; the hybridization between two Majorana modes, stemming from the spatial overlap of their wavefunctions, causes a 
splitting of their energies away from zero. The accumulated hybridization over the braid time leads to 
rotations on the Bloch sphere of the topological qubit: so-called Majorana oscillations\,\cite{chengSplittingMajoranaModes2009,chengNonadiabaticEffectsBraiding2011}. 
In other words, the larger the braid time the larger the magnitude of qubit errors. 
This effect also occurs in static systems where the MZMs hybridize too strongly, removing their topological protection and non-Abelian statistics\,\cite{chengNonadiabaticEffectsBraiding2011}.
Although Majorana modes are able to reach exactly zero energy for finely tuned systems\,\cite{leijnseParityQubitsPoor2012}, only in the thermodynamic limit 
can they exist at zero energy in the presence of perturbations. Thus, any finite system will eventually be confronted with the upper speed limit and Majorana oscillations. 

The conundrum of a topological quantum computer can hence be formulated as finding a way to satisfy the lower and upper speed limit of Majorana modes simultaneously, as given in Eq.\,\eqref{eq:speedlimit}. This will be difficult to achieve for many of the current experimental systems that claim to host Majorana modes, due to material and device constraints such as limited lengths and small superconducting gaps.

In this paper, we introduce the concept of {\it negative hybridization}, a fundamental property of Majorana modes, which can be thought of as an intrinsic error correction property of Majorana modes. We argue that negative hybridization can be used to drastically reduce $E_{\rm hyb}$ (or even set $E_{\rm hyb}\to 0$) and thus push the upper speed limit $\hbar/E_{\rm hyb}$ to larger times (or even send it to infinity). In other words, it is possible for MZMs ``without topological protection'' to have their non-Abelian statistics restored.

%
%
\textit{Negative hybridization.}---
In the following, we consider a topological superconductor hosting a Majorana bound state, i.e., a zero-energy excitation consisting of two MZMs which are spatially separated.
We can identify this excitation by analyzing the Bogoliubov--de Gennes (BdG) 
Hamiltonian in its diagonal representation, given by $H= \sum_i E_i d^\dagger_i d_i$. 
Here, the operators $d^\dagger_i$ correspond to the creation of a Boguliubov quasiparticle, which will increase the energy of the system by $E_i\geq0$. 
For the infinite system, the lowest energy quasiparticle, $d_1^\dagger$, has energy $E_1=0$. 
Furthermore, we can use these associated quasiparticle operators to construct Majorana operators:
\begin{equation}
        \gamma_{1} = d^\dagger_1 +d_1\,, \quad
         \gamma_{2} = i(d^{\dagger}_1 -d_1)\,.
    \label{eq: MZM definition}
    \end{equation}
Each of these operators is its own particle-hole conjugate, corresponding to a MZM exponentially localized at a boundary of the topological region. Together, all $\gamma_i$ satisfy a Clifford algebra. 
Exchanging two MZMs in a clockwise fashion acts on the low-energy manifold, consisting of the ground state and the occupied Majorana bound state, with the braiding operator
 \begin{equation}
     B_{12}={B_{21}}^{-1}=\rm{exp}\left(\frac{\pi}{4}\gamma_1\gamma_2\right) \ .
     \label{eq:braid}
 \end{equation}
 Critically, these braids allow for quantum Clifford gates to be enacted purely through the exchange of MZMs, forming the basis for \emph{topological quantum computation}.

Since Majorana modes in finite systems are never truly at zero energy, it is common to quantify the effect of energy splitting of the Majorana modes on the performance of the resulting gate by considering the fidelity, 
\begin{equation}
\mathcal{F}=\left|\langle\Psi_\text{target}|\Psi_\text{final}\rangle\right|^2
\label{Eq: fidelity}
\end{equation}
where $|\Psi_\text{final}\rangle = U(T)|\Psi_\text{initial}\rangle$ and the implemented braiding protocol is encoded in the time-evolution operator $U(T)$. 
By this measure, a perfect braid would have a fidelity of 1. We assert that achieving a fidelity of $\geq99$\%, exceeding the threshold required for error correction\,\cite{staceThresholdsTopologicalCodes2009,raussendorfTopologicalFaulttoleranceCluster2007}, over an extended regime of total braid times $T$, (i.e., no requirement of {\it temporal} fine-tuning) is sufficient to demonstrate non-Abelian statistics.

\begin{figure}[t!]
    \centering
    \includegraphics[width=0.98\linewidth]{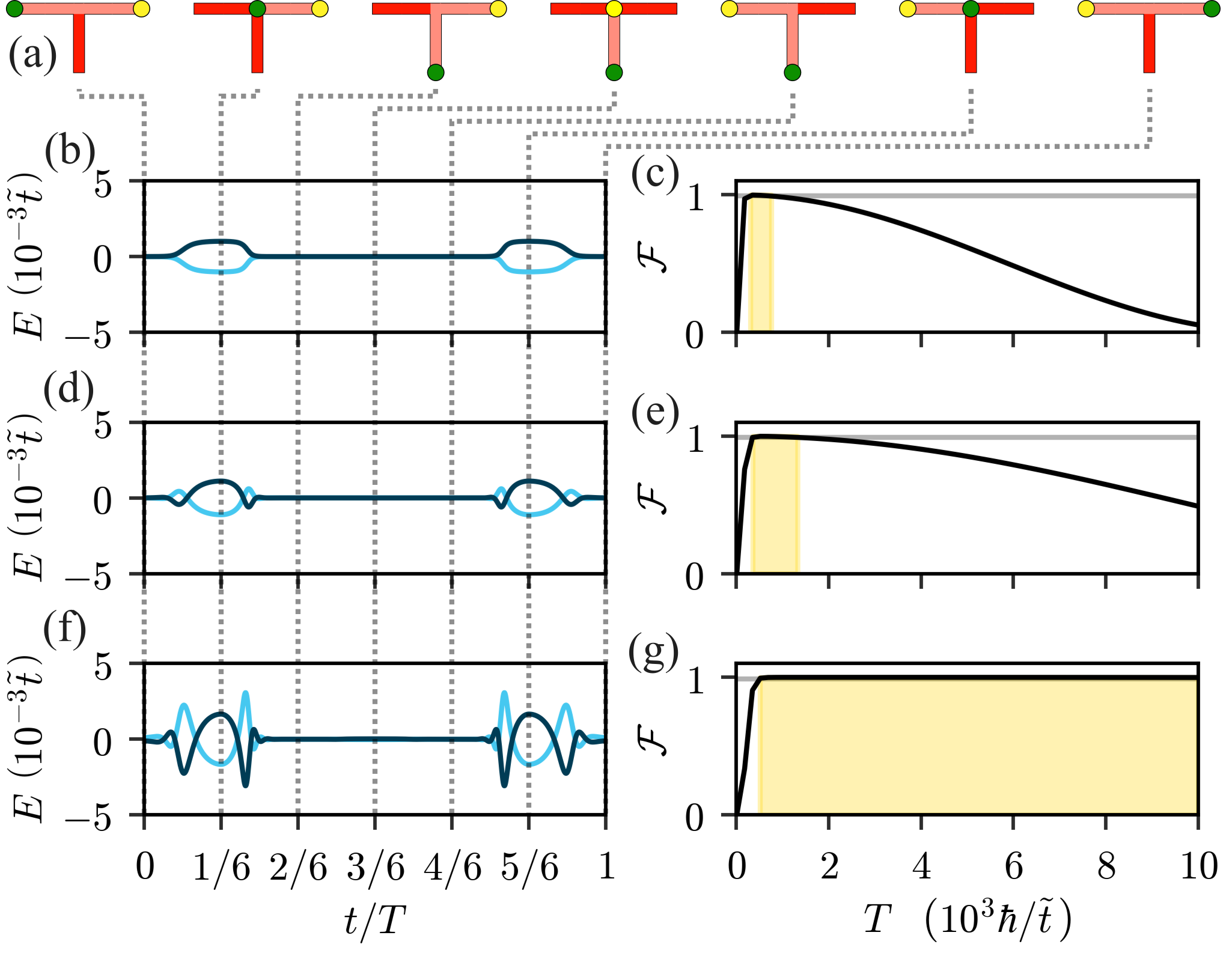}
    \caption{Energy hybridization of two MZMs performing a $\mathbf{\sqrt{\rm Z}}$ gate. (a)~Sketch of two MZMs (colored dots) being exchanged on a T-junction geometry, performing a $\sqrt{\rm Z}$ gate. Light and dark red sections correspond to topological and trivial parameters, respectively.
    (b)~Example of energy splitting between the two MZMs as a function of time $t$ (with total braid time $T$). 
    (c)~Fidelity $\mathcal{F}$ corresponding to (b) for different braid times $T$. 
    The braid times where the fidelity exceeds the threshold of 99\% to be error correctable are highlighted in yellow. 
    (d, e)~Same as (b, c) but with slightly different parameters, leading to weak negative hybridization. $\mathcal{F}$ vs.\ $T$ shows enlarged yellow region. (f, g) Same as (b, c) and (d, e) but with parameters which induce significant negative hybridization and thus vanishing $\bar{E}$. As a consequence, Majorana oscillations are absent and $\mathcal{F}$ exceeds 99\% irrespective of $T$.
}
    \label{fig:Sgate+Ebar}
\end{figure}

When MZMs are kept at a constant distance throughout a braid, 
possible for MZMs pinned to vortices in a topological superfluid or superconductor\,\cite{Ivanov2001}, their energy splitting over time 
can remain constant\,\cite{chengNonadiabaticEffectsBraiding2011}.
However, when considering other platforms such as wire networks where MZMs are physically moved (e.g.\ by means of gate-protocols), the separation, and thus energy splitting, between MZMs will change over time \cite{chengNonadiabaticEffectsBraiding2011,harper2019,bedow2024,hodgeCharacterizingDynamicHybridization2025}. 
Taking the simplest example, we consider two MZMs on a T-junction geometry performing a $\sqrt{\rm Z}$ gate\,\cite{aliceaNonAbelianStatisticsTopological2011} in which the two MZMs change their position once, as illustrated in Fig.\,\ref{fig:Sgate+Ebar}\,\textbf{a}. As expected, when MZMs get closer to each other, the energy splitting increases, see Fig.\,\ref{fig:Sgate+Ebar}\,\textbf{b} and Supplementary Movie 1 as examples\,\cite{arxiv_supplement}\nocite{Crawford2024Preparation,oregHelicalLiquidsMajorana2010a,mourikSignaturesMajoranaFermions2012}. 
The relevant quantity describing this splitting over time is the average hybridization energy:
\begin{equation}\label{Ebar}
\bar{E}=\frac{1}{T}\int_0^{T} E_{1}(t) \,dt\ .
\end{equation}
For the simplest case of two MZMs, this hybridization causes additional rotations between the MZM states, on top of the expected braid statistics, and is given by
\begin{equation}
    U_{\rm hyb}(T) = \exp\left(-\frac{\bar{E}T}{2}\gamma_1\gamma_2\right).
\end{equation}
This factor causes the resulting fidelity of an $\sqrt{\rm Z}$ gate to oscillate as $\mathcal{F}(T)=\cos^2{\left( \bar{E} T/2\right)}$\,\cite{hodgeCharacterizingDynamicHybridization2025,harper2019}; the smaller $\bar{E}$ is, the longer the fidelity remains at, or close to, one.

The spectra shown in Fig.\,\ref{fig:Sgate+Ebar}\,(b,\,d,\,f) are obtained by diagonalizing the BdG Hamiltonian where, by construction, the doubling of the degrees of freedom leads to identical positive and negative energies. 
Typically, one would only consider the positive half of such spectra as physical. Yet, when inspecting spectra similar to the examples in Fig.\,\ref{fig:Sgate+Ebar}\,(d, f), it is clear that the energy of the MBS crosses through $E=0$ and spans both positive and negative values.

This energy crossing corresponds to a swapping of the ground state parity, where the 
original many-body ground state becomes higher in energy than the first excited state. Since the two states differ by one quasi-particle, the parity 
must change (see Supplement\,\cite{arxiv_supplement}). 
These parity swaps can be calculated by taking the sign of the Pfaffian of the BdG Hamiltonian in the Majorana basis\,\cite{kitaevUnpairedMajoranaFermions2001}. 
This confirms the zero-energy crossings in 
Fig.\,\ref{fig:Sgate+Ebar}\,(d,\,f).

As a consequence of this effect,
when integrating the energy splitting to calculate $\bar{E}$, 
hybridization of the positive sections will cancel with the negative sections, thereby reducing the magnitude of $\bar{E}$. This can be seen in Fig.\,\ref{fig:Sgate+Ebar}\,(b,\,d,\,f) where their corresponding fidelities, Fig.\,\ref{fig:Sgate+Ebar}\,(c,\,e,\,g), exceed the fault tolerance threshold for progressively wider ranges of braid times due to the reduced hybridization error. 
We refer to the phenomenon of  $E_{1}(t)$ changing sign, leading to an improvement in braid fidelity, as {\it negative hybridization}.
Due to the presence of negative hybridization, the requirement that a ``good'' braid has $\bar{E}=0$, does not imply that the $E_1$ must remain at zero for the entire process, suggesting that accurate gates may be achievable even in systems where hybridization is unavoidable.

The remainder of this paper will explore ways in which the braid protocol may be adapted to induce negative hybridization, thus improving quantum gates afflicted with fidelities far below the required accuracy $(<99\%)$ to far above $(>99\%)$, thus restoring fault tolerance\,\cite{raussendorfTopologicalFaulttoleranceCluster2007,staceThresholdsTopologicalCodes2009}.

%
%
\textit{Negative hybridization enforced by a local gate.}---
In the following, we will investigate a simple scenario for the utilization of negative hybridization which is potentially controllable and implementable in various platforms. To this end, we study the simplest topological superconductor model, the Kitaev chain. It is governed by the Hamiltonian
\begin{equation}
    H_{\rm K}(t) = \sum_{j=0}^N -\mu_j(t)c_j^\dagger c_j^\pd + \sum_{j=0}^{N-1} -\tilde{t}\,c_{j+1}^\dagger c_j^\pd + \Delta c^\dagger_{j+1} c_j^\dagger +{\rm H.c.}\,,
    \label{Eq: Kitaev Chain Hamiltonian}
\end{equation}
where $\mu_j(t)$ is a site- and time-dependent chemical potential, $\tilde{t}$ is the hopping strength, and $\Delta$ is the superconducting order parameter. MZMs can be moved by carefully changing the chemical potentials of certain sites between a topological ($ |\mu/\tilde{t}| < 2$) and a trivial ($|\mu/\tilde{t}|>2$) value, which we refer to as $\mu_\text{topo}$ and $\mu_\text{triv}$. For details about the ramping protocol and how to extend the Kitaev chain to a 
T-junction which allow braids, see the End Matter.

\begin{figure}[t!]
    \centering
\includegraphics[width=0.98\linewidth]{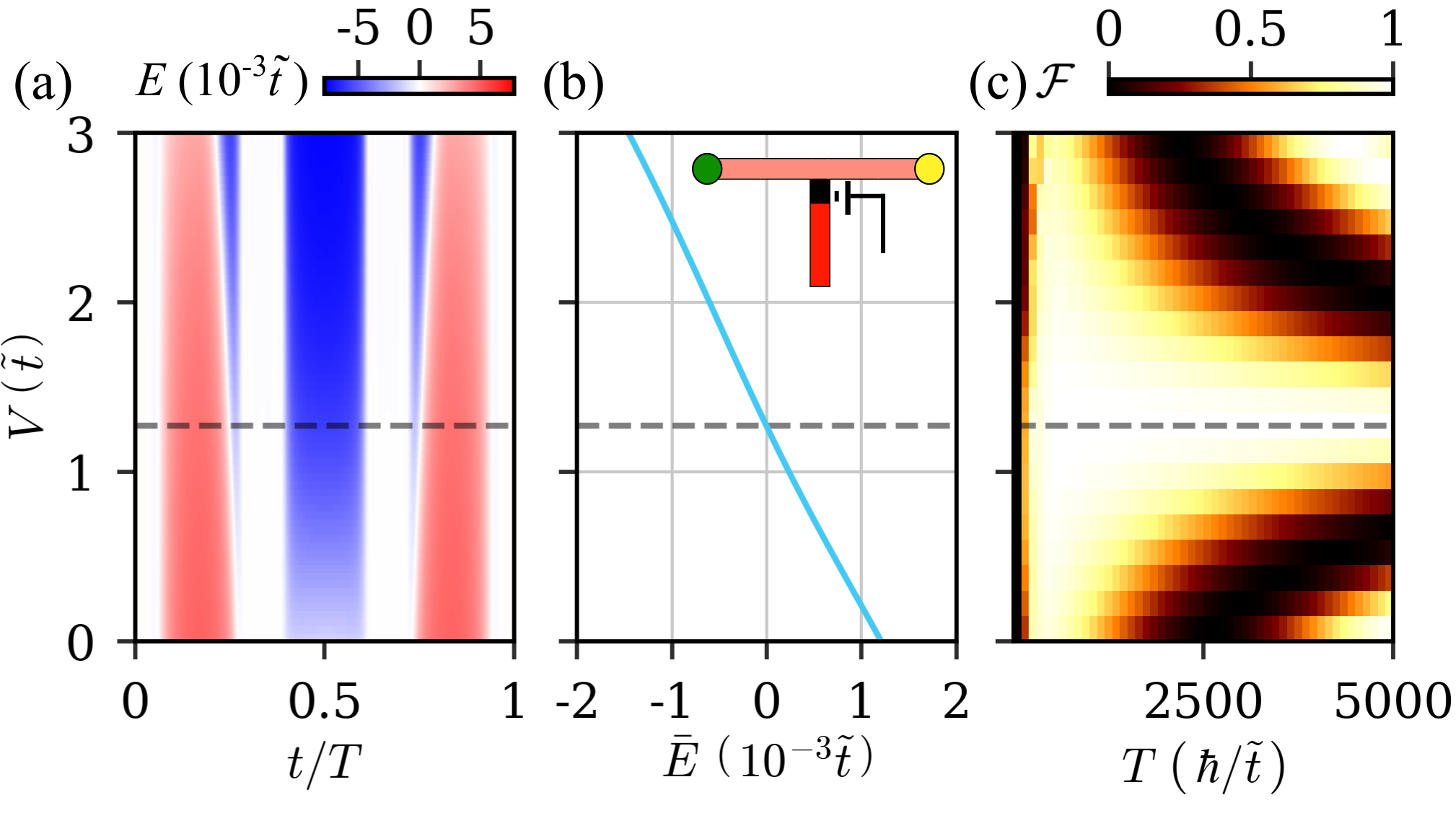}
    \caption{Negative hybridization induced by a local gate potential. 
    (a) The hybridization energy between the MZMs throughout the braid, for varying applied voltage strengths, $V$, located at the site below the junction. As $V$ increases, the main parity swaps occur earlier in the braid, leading to greater amounts of negative hybridization.
    (b) Integrated energy, $\bar{E}$, for the braid as a function of $V$. The dashed line at $V=1.27\tilde{t}$ marks the potential where the braid has $\bar{E}=0$. Inset: T-junction with the additional local gate highlighted in black.
    (c) Fidelity of the braid for various $V$ and $T$. 
    Around $V=1.27\tilde{t}$ the fidelity remains close to 1, even for large values of $T$.
    }
    \label{fig:alt_site23}
\end{figure}

We consider a T-junction geometry where an additional local gate-potential with strength $V$, satisfying $\mu_\text{topo}<V<\mu_\text{triv}$, is applied to the top site of the vertical leg (see inset of Fig.\,\ref{fig:alt_site23}\,(b)). For simplicity, we restrict this gate to a single site (for extensions to more than a single site\,\cite{aliceaNonAbelianStatisticsTopological2011,bauer2018}, see Supplement\,\cite{arxiv_supplement}), modifying $\mu$ locally, i.e., $\mu_{\rm site}(t) \to \mu_{\rm site}(t)-V$.

At the start of the braid the site has chemical potential $\mu_{\rm site}(0) = \mu_\text{triv}-V>0$, but during the braid there are times when the bottom leg becomes topological and $\mu_{\rm site}(t) = \mu_\text{topo}-V<0$, changing sign. Due to the chosen location of the potential, this change in sign occurs when the MZMs are 
strongly hybridizing. 
In the limit $|\Delta|=|\tilde{t}|$, the parity 
simply becomes the product of the chemical potentials on all sites of the chain. Despite the number and timing of parity swaps changing as the magnitude of $\Delta$ is lowered, it remains a valid approximation, as shown in the Supplement\,\cite{arxiv_supplement}.
Therefore, if both $\mu_\text{topo}$ and $\mu_\text{triv}$ have the same sign, the parity of the system remains constant throughout a braid. 
However, by changing the sign of the chemical potential on just one site, we are able to induce negative hybridization.

As seen in Fig.\ref{fig:alt_site23}\,(a), where the hybridization energy throughout a $\sqrt{Z}$ gate is plotted as a function of time and gating potential, increasing $V$ leads to an earlier parity swap as $\mu_{\rm site}$ is brought closer to zero, 
thus introducing more negative hybridization. 
From the hybridization energy of each braid, we obtain $\bar{E}(V)$, shown in Fig.\,\ref{fig:alt_site23}\,(b). As expected, increasing the amount of negative hybridization by increasing $V$ leads to a reduction of $\bar{E}$, with $\bar{E}=0$ for $V\approx1.27\tilde{t}$. In Fig.\,\ref{fig:alt_site23}\,(c) the fidelity of the braid is plotted for varying braid times, $T$. 
These show the expected Majorana oscillations for the clean system, limiting the range of $T$ for successful braids, but as $V$ approaches $1.27\tilde{t}$, the region where $\mathcal{F}>0.99$ rapidly extends to large times. At this voltage, the fidelity is essentially independent of braid time, leaving only the contributions from the braiding statistics, thus raising the gate fidelity well above the fault tolerance threshold.

It should be noted that the strength of the potential itself modifies the magnitude of MZM hybridization\,\cite{peetersEffectImpuritiesDisorder2024}. 
Therefore, the value of $V$ that minimizes $\bar{E}$ is difficult to determine \textit{a priori}.
However, calibration through 
repeating the same braid for changing values of $V$, seems to be a rather simple task, and as demonstrated by Fig.\,\ref{fig:alt_site23}, even a non-perfectly calibrated gate potential can significantly enhance the braiding performance. 
We stress that for the quantum dot platform\,\cite{leijnseParityQubitsPoor2012,Liu2022,bordin2025} which hosts ``poor man's Majoranas'', our described protocol for utilizing negative hybridization could be directly implemented.

%
%
\textit{Negative hybridization due to symmetric braids.}---
The above reduction of $\bar{E}$ requires the calibration of a 
system parameter in order to significantly reduce 
the dynamic phase accumulated over the braiding process, with perfect phase cancellation necessitating fine-tuning. 
 Thus, for our second example, we use the original braid as the basis of the new protocol, where the second half of the braid is built to have the same magnitude of hybridization but opposite sign, 
leading to $\bar{E}=0$. 
The key is to utilize the symmetry properties of topological superconductors, here specifically the Kitaev model. 
Similar reasoning has been proposed, where $X$ gates are acted on ancillary modes to induce parity swaps\,\cite{karzigUniversalGeometricPath2016,martinDoubleBraidingMajoranas2020}, however, our method relies on the intrinsic parity switching of the MZMs and thus does not rely on external gates or zero modes.

The antisymmetry of the spectrum of the Kitaev chain for odd chain lengths\,\cite{hegdeMajoranaWavefunctionOscillations2016,leumerExactEigenvectorsEigenvalues2020} with respect to positive and negative chemical potentials allows us to perform identical operations but with opposite parity. This method can be extended to other spinful topological superconductor models\,\cite{setiawanTransportSuperconductorNormal2017,oreg-10prl177002,lutchynMajoranaFermionsTopological2010} as shown in the Supplement\,\cite{arxiv_supplement}. We stress that restricting segments of a T-junction to an odd number of lattice sites or atoms is unproblematic for quantum dot arrays\,\cite{Liu2022,bordin2025} and for magnet-superconductor hybrid structures, where atomic chains are built atom-by-atom\,\cite{kim2018,schneider21,rachel-25pr1}.

\begin{figure}[t!]
    \centering
    \includegraphics[width=0.98\linewidth]{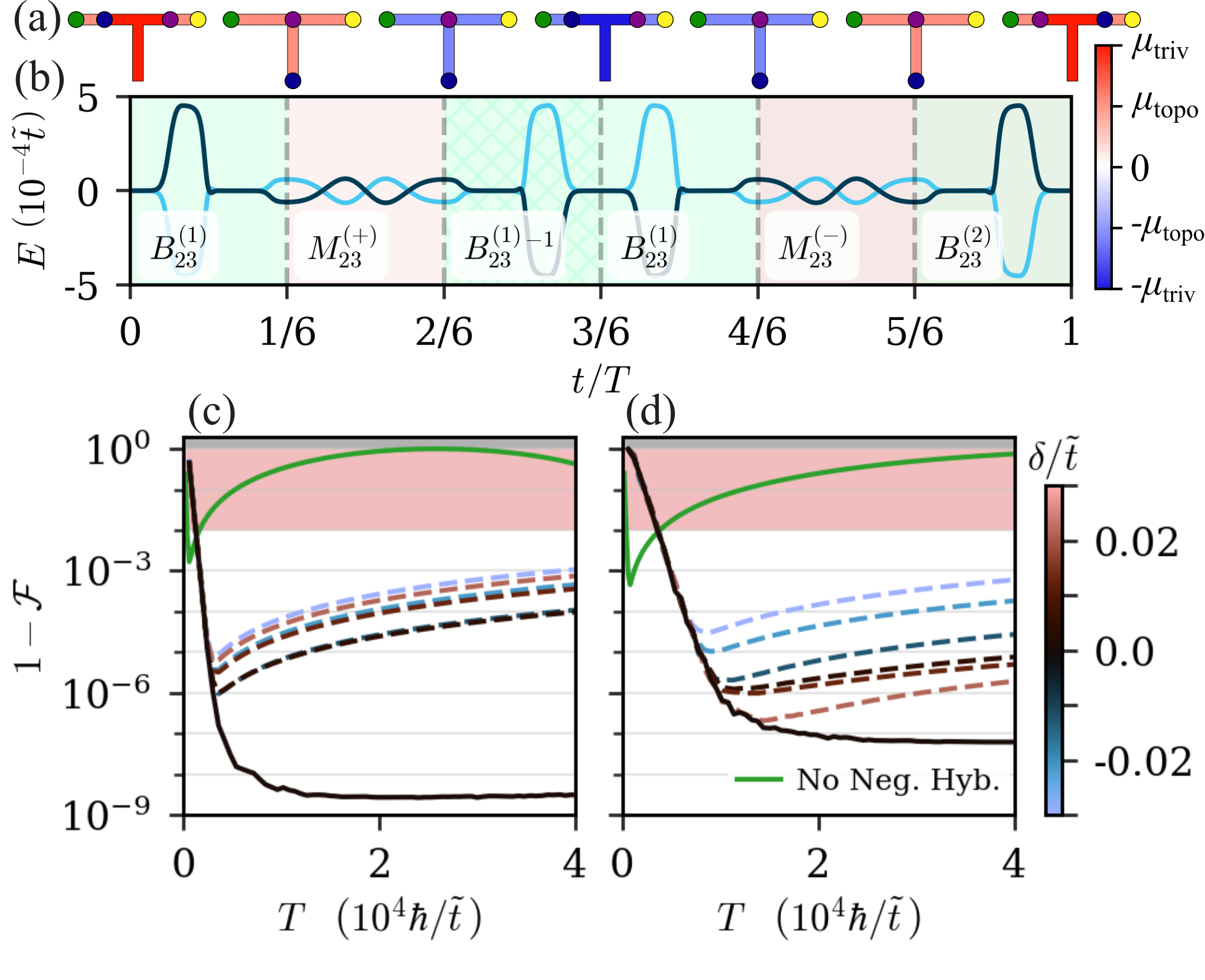}
    \caption{Symmetric braids: errors of corrected and uncorrected single-qubit gates. 
    (a) Schematic of the corrected $\sqrt{X}$ gate showing MZMs (colored) at the borders of the topological (light red and light blue) and trivial (dark red and dark blue) regions. The red regions correspond to positive chemical potentials, while the blue ones indicate that the chemical potential is negative.
    (b) Energy spectrum of the Majorana bound states for the corrected $\sqrt{X}$ gate as function of time, labeled with the braid operations taking place.
    (c) Error in fidelity for $\sqrt{X}$ gate vs.\ braid times. The red region covers the fidelities which exceed the 1\% error threshold. The green curve is the uncorrected result, all other lines have been corrected by means of symmetric braids. $\delta=0$ (solid black) is perfectly symmetric, and finite $\delta$ values (dashed) indicate deviation from perfect symmetry. (d) Same as (c) but for the $\sqrt{Z}$ gate.
    }
    \label{fig:basic_gate_fidelities}
\end{figure}  

For the proposed method to work, two challenges need to be overcome: (1) The additional operations must not contribute to the braiding statistics. 
(2) The process which induces the parity swap \emph{cannot} create uncanceled hybridization itself.

To produce a trivial movement of the MZMs which accumulates the same hybridization as a braid, yet acts as an identity gate, we first decompose the original braid into two halves: $B_{ij}^{(1)}$ and $B_{ij}^{(2)}$. For illustration purposes, suppose that the first operation corresponds to the process $t=0$ to $t=3/6T$, depicted in Fig.\,\ref{fig:Sgate+Ebar}\,(a), while the second operation to the process $t=3/6T$ to $t=T$. 
The original braid is therefore defined as $B_{ij}^{(2)} B_{ij}^{(1)}$, while the new trivial operations are given by either $B_{ij}^{(1)} {B_{ij}^{(1)}}^{-1}$ or ${B_{ij}^{(2)}} {B_{ij}^{(2)}}^{-1}$.

Fig.\,\ref{fig:basic_gate_fidelities}\,(a) depicts the state of the system as a corrected $\sqrt{X}$ gate is performed, where the dots represent MZMs, and the color of each section denotes the chemical potential as given by the color bar. 
The time-dependent hybridization energy is plotted in Fig.\,\ref{fig:basic_gate_fidelities}\,(b), where each sixth of the braid is colored and labeled according to the process occurring. 
At the midpoint of the uncorrected braid, $T=1/6$, all sites are topological with $\mu=\mu_\text{topo}$ and the entire system can be simultaneously ramped from $\mu_\text{topo}$ to $-\mu_\text{topo}$, defining the operation $M^{(+)}_{ij}$ whose counterpart, $M^{(-)}_{ij}$, transforms the system back from a negative chemical potential to a positive one.
As the $M^{(\pm)}_{ij}$ protocols are symmetric in $\mu$, they have no net effect on $\bar{E}$, as seen in Fig.\,\ref{fig:basic_gate_fidelities}\,(b).

We note that swapping the chemical potentials for a braid 
in a $\sqrt{Z}$ gate is more difficult as
there are always regions of the junction which are trivial. These cannot directly go from $\mu_\text{triv}$ to $-\mu_\text{triv}$ without causing the formation of new MZMs --- effectively introducing quasiparticle poisoning. To avoid this, we employ a more sophisticated procedure, as outlined in the Supplement\,\cite{arxiv_supplement}.

By taking advantage of the parity's antisymmetric structure with respect to the chemical potential, the routines for the corrected braids are agnostic to the specific parameters of the system, thus we do not need any prior knowledge of the hybridization dynamics to produce a braid with no net hybridization.


To demonstrate the effectiveness of these protocols we simulate the corrected braids corresponding to $\sqrt{X}$ and $\sqrt{Z}$ gates. 
The resulting error in the fidelity, $1-\mathcal{F}$, as a function of total braiding time, $T$, is shown in Fig.\ref{fig:basic_gate_fidelities}\,(c,\,d), compared to the error of the conventional, {\it uncorrected} gates. The red region shows errors larger than 1\% where quantum computing fails. 
The uncorrected (solid green) curve briefly escapes the red region, but then returns due to the Majorana oscillations. 
Meanwhile, the corrected (solid black) curve decreases monotonically with the diabatic error before plateauing at an exceptionally small error $1-\mathcal{F}<10^{-7}$, far below the level required for error correction.

As it is unrealistic to assume arbitrarily precise control over the braid, we introduce the parameter $\delta=\mu^+_\text{topo} + \mu^-_\text{topo}$, which quantifies the asymmetry between the positive and negative chemical potentials. 
The color of the dashed lines in Fig.\,\ref{fig:basic_gate_fidelities}\,(c,\,d) correspond to specific values of $\delta$. In general we see that although this imperfection reintroduces hybridization dependent error, it remains orders of magnitude lower than the uncorrected case, even when $\delta$ is a considerable fraction of $\mu_\text{topo}$, highlighting the robustness of the corrected gates. 


\textit{Outlook.}---
The importance of our results is merely a reflection of the fundamental property of the Majorana modes: regardless of what model or what specific platform considered, any Majorana mode can be used for braiding and topological quantum computing by diminishing the hybridization error and thus pushing  the upper speed limit to large times. Already today, one could easily include negative hybridization for reducing braiding errors in quantum dot based few-site Kitaev chains\,\cite{Liu2022,bordin2025}.
We envision that future topological superconductor platforms can be tailored to directly implement negative hybridization or at least make its usage easily accessible. Our work thus suggests that a Majorana-based quantum computer could in principle be improved to reach fault-tolerance by exploiting negative hybridization.

\textit{Acknowledgments.}---We thank E. Mascot for helpful discussions.
S.R.\ acknowledges support from the Australian Research Council through Grant No.\ DP200101118 and DP240100168.
This research was undertaken with the assistance of resources from the National Computational Infrastructure (NCI Australia), an NCRIS enabled capability supported by the Australian Government.
This research was supported by The University of Melbourne’s Research Computing Services and the Petascale Campus Initiative.

\textit{Data availability.}---The data that support the findings of this article are openly available\,\cite{dataset}.

\section*{End Matter}
\textit{CNOT gate corrected via negative hybridization.}---Individual corrected gates can be easily combined and extended beyond a single qubit. To this end, we demonstrate the CNOT gate on a 2-qubit system, comprised of 6 MZMs. We note that the further extension to more than two qubits is straight-forward e.g.\ via projective measurements\,\cite{Frey2025,Hodge2025proj}, where it is sufficient to correct single and two-qubit gates.

\begin{figure}
    \centering
    \includegraphics[width=0.98\linewidth]{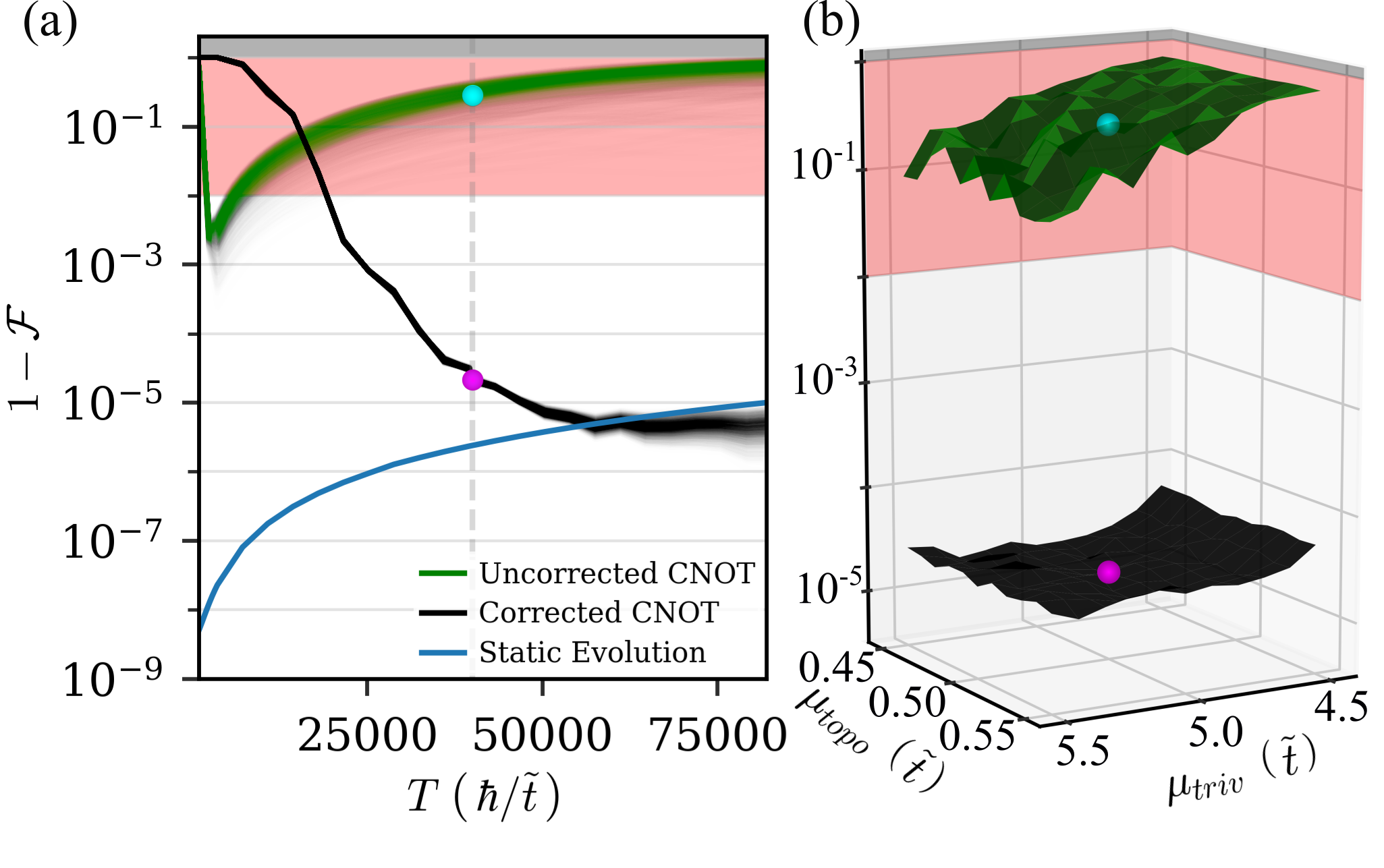}
    \caption{CNOT gate error. (a) Loss of fidelity for the corrected and uncorrected CNOT gates vs.\ braid time, compared to the natural time evolution of the system. Results slightly depend on initial states, leading to the spread of the uncorrected curve.
    (b) Average fidelities for the corrected and uncorrected gates over a range of chemical potential values, demonstrating the robustness of the protocol. Results are for the braid time $T=4000\hbar/\tilde{t}$, indicated by the dashed line in (a) The dots indicate common data between (a) and (b).
    }
    \label{fig:CX gate}
\end{figure}

The CNOT gate in the even parity subspace can be expressed as $B_{12}B_{23}B_{12}B_{65}B_{34}B_{21}B_{12}B_{23}B_{12}$\,\cite{mascotManyBodyMajoranaBraiding2023,Georgiev-2008NuclPhysB789}. Therefore, the corrected CNOT gate is simply a collection of multiple braids, each being individually corrected. In Fig.\,\ref{fig:CX gate}\,(a) we show the braiding error, $1-\mathcal{F}$, of the CNOT gate. As the fidelity of the CNOT gate depends on the initial state, we test multiple random initial states, shown green for the uncorrected and black for the corrected gates. The improvement due to negative hybridization is striking, showing that the proposed method to produce corrected braids can indeed be extended and applied to complex circuits with multiple qubits. The blue line in Fig.\,\ref{fig:CX gate}\,(a) is the error of the static system, i.e., the accumulated error from the hybridization of MZMs simply waiting in their initial positions. Although much smaller than the error during the braids where the MZMs move closer to each other, it is neglected in the correction process and thus forms a lower bound on the error of the corrected gates.
In Fig.\,\ref{fig:CX gate}\,(b) we show deviations from the parameters $\mu_{\rm topo}$ and $\mu_{\rm triv}$ used in Fig.\,\ref{fig:CX gate}\,(a) at time $T=4000\hbar/\tilde{t}$, indicated by the dots and dashed line. This demonstrates the robustness of the protocol with respect to different choices of parameters. The errors of corrected and uncorrected CNOT gate remain separated by roughly four magnitudes for the wide range of chemical potentials examined. 

\textit{Method.}---
\begin{figure*}[t!]
    \centering
    \includegraphics[width=0.98\linewidth]{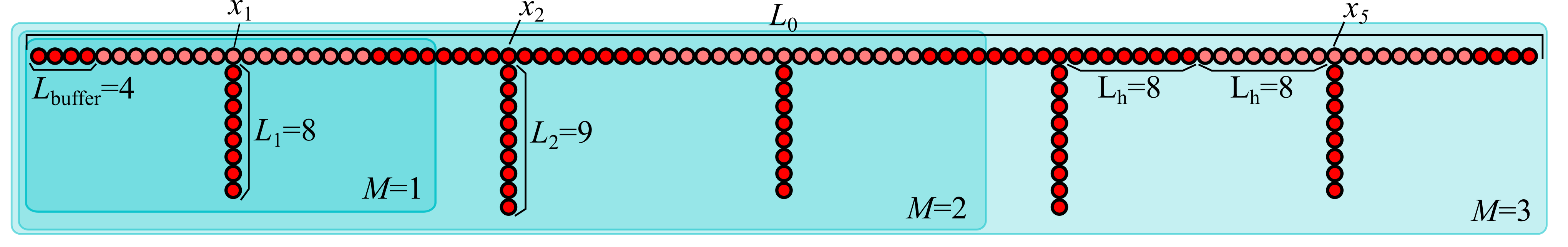}
    \caption{Schematic of extended T-junction system. The sites are represented by dots, with adjacent ones being connected by hopping and superconducting terms. The color denotes the chemical potential at the site in the initial configuration, where light red is $\mu_{\rm topo}$ and dark red is $\mu_{\rm triv}$. The sites enclosed within each box make up a T-junction, 3T-junction and 5T-junction for $M=1$, $M=2$ and $M=3$, respectively. The horizontal leg had $L_0$ sites, while the vertical legs have $L_i$, with the $i^{th}$ vertical leg attaches to the horizontal backbone at site $x_i$. The system has four buffer sites at each end, labeled by $l_{\rm buffer}$, while the spacing between MZMs on the horizontal leg is given by $L_h$ as shown.}
    \label{fig:5T-junction}
\end{figure*}
All of the simulations were performed on Kitaev chains which are combined into T-junction geometries. The Hamiltonian for a single Kitaev chain with $N$ sites is given by  Eq.\,\eqref{Eq: Kitaev Chain Hamiltonian} in the main text. Its extension to a multi T-junction geometry, consisting of multiple vertical legs connected to a horizontal chain (see e.g.\ Fig.\,\ref{fig:5T-junction}), is given by
\begin{equation}
    \begin{aligned}
    H(t) = \sum_{i=0}^{2M-1} & \left[\sum_{j=1}^{L_i} -\mu_{i,j}(t)c^\dagger_{i,j} c_{i,j} \right. \\
    +& \left.\sum_{j=1}^{L_i-1} \left(-\Tilde{t}c^\dagger_{i,j} c_{i,j+1} + \Delta e^{i\phi_i} c^\dagger_{i,j} c^\dagger_{i,j+1} + {\rm H.c.}\right)\right]\\
    +&  \sum_{i=1}^{2M-1} \left(-\Tilde{t}c^\dagger_{0,x_i} c_{i,1} + \Delta e^{i\phi_i} c^\dagger_{0,x_i} c^\dagger_{i,1} + {\rm H.c.}\right)\ .
    \end{aligned}
    \label{full_ham}
\end{equation}
Here $M$ is the number of MZM pairs and $L_i$ is the number of sites in the $i^{\rm th}$ leg, ordered such that the horizontal backbone is $L_0$ and the vertical legs, from left to right, are $L_1$ through $L_{2M-1}$. As the MZMs are evenly spaced along the horizontal chain, it is convenient to use half that distance, $L_h$, and the number of buffer sites, $L_{\rm buffer}$, to express the horizontal length: $L_0 = 2L_{\rm buffer}+(2M-1)(2L_h+1)$. Each pair of adjacent MZMs has a vertical leg between them so that they can be braided without the movement of any of the other MZMs. These are attached to the horizontal leg at sites $x_i=L_{\rm buffer}+(2i-1)L_h+i$. To facilitate the parity swaps required for the symmetric braids, the lengths of the vertical legs alternate such that for $n\in\mathbb{Z}^+$, $L_{2n-1}=L_v$ and $L_{2n}=L_v+1$ where $L_v$ is the length of the odd numbered legs. These features are labeled for the example shown in Fig.\,\ref{fig:5T-junction}, and the purpose of the buffer sites is outlined in the Supplement\,\cite{arxiv_supplement}.

We define the superconducting phases of each section as
\begin{equation}
    \phi_i = 
    \begin{cases}
        0 & i=0,\\
        \frac{-\pi}{2} & i>0,
    \end{cases}
\end{equation}
to mirror the $p_x+ip_y$ superconductivity that provides the correct braiding statistics, as all of the vertical legs ($i>0$) are orthogonal to the horizontal backbone ($i=0$).

As mentioned in the main text, this Hamiltonian can be diagonalized and expressed as $H= \sum_i E_i d^\dagger_i d_i$ where the Bogoliubov quasiparticles, $d^\dagger_i$, are the energy eigenstates of the system.

To move the MZMs around, we ramp the local chemical potentials over time, using a smooth-step function to modulate the chemical potential, defined as
\begin{equation}
    s(q) = \begin{cases}
        0 & q<0, \\
        q^2(3-2q) & 0\leq q\leq1, \\
        1 & q>1,
    \end{cases}
\end{equation}
where $q=0$ ($q=1$) corresponds to the beginning (end) of the process. By staggering these on-site modulations along a number of adjacent sites, we can generate a smooth ramp that is able to move the MZMs through each section of the braids with minimal loss. The chemical potential for the $n^{th}$ site in such a chain is given by
\begin{equation}
    \mu_n(\tau)=\mu_{\rm triv} + (\mu_{\rm topo}-\mu_{\rm triv}) s\left(\tau(1+\alpha(N-1))-\alpha n \right)
\end{equation}
where $0\leq\tau\leq1$ is the fraction of the ramp that has been completed. The total number of sites involved is $N$, and $\alpha$ is the delay coefficient.  
The choice $\alpha=0$ ramps all of the sites simultaneously, while $\alpha=1$ ramps each site completely sequentially. Unless otherwise specified, all results shown use a delay coefficient of $\alpha=0.025$.
In the case where static disorder is considered, we add to each chemical potential a constant random offset $\mu_i(t)$ , where $\mu_i(t)\to\mu_i(t)+\delta\mu_i$.

\begin{table}[h!]
    \centering
    \begin{tabular}{c||c|c|c|c|c|c|c}
    System     & $M$ & $L_h$ & $L_v$ & $L_{\rm buffer}$ & $\Delta$ & $\mu_{\rm topo}$ & $\mu_{\rm triv}$\\
    \hline
     Fig.\,\ref{fig:alt_site23} & 1 & 5 & 11 & 0 & $0.7\tilde{t}$ & $0.5\tilde{t}$ & $5\tilde{t}$ \\
     Fig.\,\ref{fig:basic_gate_fidelities}(b, c) & 2 & 8 & 8 & 0 & $0.8\tilde{t}$ & $0.5\tilde{t}$ & $5\tilde{t}$ \\
     Fig.\,\ref{fig:basic_gate_fidelities}(d) & 1 & 8 & 8 & 0 & $0.8\tilde{t}$ & $0.5\tilde{t}$ & $5\tilde{t}$\\
     Fig.\,\ref{fig:CX gate}(a) & 3 & 8 & 8 & 4 & $0.8\tilde{t}$ & $0.5\tilde{t}$ & $5\tilde{t}$ \\
     Fig.\,\ref{fig:CX gate}(b) & 3 & 8 & 8 & 4 & $0.8\tilde{t}$ & As shown & As shown  
    \end{tabular}
    \caption{System parameters. The values of the parameters referenced in the Hamiltonian given by Eq.\,\eqref{full_ham} used to construct the systems simulated in each figure. For all simulations, unless specified, a delay coefficient of $\alpha=0.025$ is used.}
    \label{tab:parameters}
\end{table}

The parameters used to obtain the results shown in each figure are summarized in Table \ref{tab:parameters}.

The braiding routines for each uncorrected gate are the typical protocols as defined previously\,\cite{mascotManyBodyMajoranaBraiding2023}. Specific details for the corrections to the protocol introduced in the section examining symmetric braids can be found in the Supplement, and are illustrated in the accompanying Supplementary Movies 2 and 3\,\cite{arxiv_supplement}.

By diagonalizing the initial Hamiltonian in Eq.\,\eqref{full_ham}, $H(0)$, we obtain the single particle excitation eigenstates of the system which we can time evolve over the braid using the time evolution operator
\begin{equation}
    U(T) = \mathcal{T}\exp\left[-\frac{i}{\hbar}\int_0^T dt H(t)\right].
\end{equation}
This is numerically approximated as a product of infinitesimal time steps using a $4^{th}$ order Runge-Kutta routine (as discussed and bench-marked in Ref.\,\onlinecite{mascotManyBodyMajoranaBraiding2023}).

The time evolution operator is used to define the time dependent ground state, $|\mathbf{0}_d(t)\rangle$, and quasiparticle operators,  $d^\dagger(t)$, which through the use of the  time dependent Pfaffian (TDP) method\,\cite{mascotManyBodyMajoranaBraiding2023}, can be used to construct the many-body states $|\mathbf{n}(t)\rangle=\prod^N_{i=1}(d^\dagger_i(t))^{n_i}|\mathbf{0}_d(t)\rangle$, where, for a system hosting $N$ Majorana bound states, the vector $\mathbf{n}=(n_1,\,n_2,\,...,\, n_N)$ indexes whether the $i$th Majorana bound state is occupied $(n_i=1)$ or empty $(n_i=0)$. Overlaps between initial and final states, of the form
\begin{equation}
    \left\langle \mathbf{n}(T)|\mathbf{n}'(0)\right\rangle,
\end{equation}
are then used to determine the appropriate fidelity of each gate.

\bibliography{Neghyb}

\end{document}


\newcommand{\FigOne}{1}
\newcommand{\FigTwo}{2}
\newcommand{\FigThree}{3}
\newcommand{\FigFour}{4}

\newcommand{\MzmDefinition}{(2)}
\newcommand{\BraidEquation}{(3)}
\newcommand{\EqEbar}{(5)}
\newcommand{\KitaevChainHamiltonian}{(7)}
\newcommand{\EqFidelity}{(4)}
\newcommand{\FullHam}{(8)}
\newcommand{\RampEq}{(11)}

\newcommand{\CnotSec}{IV}

\title{{\it Supplementary Material:}\\[5pt]
Negative Hybridization:\\
a Potential Cure for Braiding with Imperfect Majorana Modes}

\author{Cole Peeters}
\affiliation{School of Physics, University of Melbourne, Parkville, VIC 3010, Australia}
\author{Themba Hodge}
\affiliation{School of Physics, University of Melbourne, Parkville, VIC 3010, Australia}
\author{Stephan Rachel}
\affiliation{School of Physics, University of Melbourne, Parkville, VIC 3010, Australia}

\maketitle

\tableofcontents



\section{Extension to Systems with Multiple Majorana Zero Modes}
 
For systems with more than one Majorana bound state, the definition given by Eq.\,\MzmDefinition\ in the main text must be extended. In general,
\begin{equation}
        \gamma_{2i-1} = d_i^\dagger +d_i\,, \quad
         \gamma_{2i} = i(d^{\dagger}_i -d_i)\,,
    \label{MZM definition}
\end{equation}
where $d_i$ ($d_i^\dagger$) is the annihilation (creation) operator for the $i\in\{1,...,N\}$ lowest energy excitation of the Hamiltonian $H= \sum_i E_i d^\dagger_i d_i$, and $N$ is the number of pairs of Majorana zero modes (MZMs) present.

It is necessary to consider these larger systems because multiple Majorana bound states are required to construct a single qubit. 
At first glance it may seem that one Majorana bound state is sufficient, as one can define $|0\rangle_{\rm log}$ to be the ground state and $|1\rangle_{\rm log}$ to be the state with only the Majorana bound state occupied, giving the required logical basis states. However, these states are in different parity sectors, and therefore cannot be transformed into each other as would be required to perform, for example, the $X$ gate.

The two independent parity sectors arise from the superconductivity terms ($\Delta c_i c_{i+1}+\Delta^* c^\dagger_{i+1}c^\dagger_i$), which, despite breaking particle number conservation, still conserve fermionic parity, as the addition or removal of each Cooper pair always involves two electrons. 
In order to ensure that all physical states have the same parity, the following form for the logical states is chosen
    \begin{equation}
        \quad \quad\quad|0(t)\rangle_{\rm log} = |\mathbf{0}_d(t)\rangle\,, \quad
        |1(t)\rangle_{\rm log} = d^\dagger_1(t)d^\dagger_2(t)|\textbf{0}_d(t)\rangle\,,
        \label{logical basis}
    \end{equation}
where $\ket{\textbf{0}_d(t)}=U(t)\ket{\textbf{0}_d(0)}$ is the quasiparticle vacuum at time $t$, defined such that $d_i(0)\ket{\textbf{0}_d(0)}=0, \; \forall \, i$. 
The time-evolved quasiparticle operator, $d_i(t)$, is given by $d_i(t)=U(t)d_i(0)U^\dagger(t)$, where $U(t)$ is the time evolution operator which evolves the initial states to time $t$. 

We can use this definition of a logical qubit to straight-forwardly construct multi-qubit systems, where each qubit consists of four MZMs. 
This is usually referred to as `sparse' encoding, and allows for the implementation of single qubit Clifford gates by braiding within each qubit, however, offers no way to perform entangling gates.

An alternative to this is the `dense' encoding, which constructs $N$ logical
 qubits out of $2N+2$, rather than $4N$ MZMs. 
 This is possible as the total parity of the system is fixed by the final pair of MZMs, acting as an ancilla qubit. 
 Although dense encoding enables the implementation of entangling gates, there is no scalable way to universally perform single qubit gates beyond two qubits \cite{Frey2025,Hodge2025proj}. 
While beyond the scope of this paper, we stress a universal gate set for an \emph{arbitrary} number of qubits is enabled by utilizing \emph{projective measurements}, which enables mapping between one encoding to another. 
It follows that, for the purpose of topological quantum computation, it is sufficient to consider the fidelity of each individual braid, as any desired Clifford gate can be implemented by the composition of individual braiding operations.

Thus, in the main paper, we investigate the stability of the $\sqrt{X}$ and $\sqrt{Z}$ gates, implemented via individual braiding operations as given in Fig.\,\FigTwo\ and \FigThree, and further, we utilize the dense encoding to test the stability of the CNOT gate where six MZMs are used to represent two qubits as shown in Fig.\,\FigFour.

For the single qubit gates, the sparse and dense encodings are indistinguishable, as they both require four MZMs.
Demonstrating the efficacy of these gates forms the building blocks of a topological quantum computer, as all gates on larger systems may be decomposed to these basic operations.
Any arbitrary braiding operation between two MZMs, $\gamma_i$ and $\gamma_j$, is represented by the \emph{braiding operator}, $B_{ij}$, which is encoded in the time-evolution when the MZMs are spatially exchanged. 
This is generalized from the two MZM operator $B_{12}$, explicitly defined in Eq.\,\BraidEquation\ in the main text, and is given by
\begin{equation}
    B_{ij}={B_{ji}}^{-1}=\rm{exp}\left(\frac{\pi}{4}\gamma_i\gamma_j\right)\ . 
    \label{eq:braid}
\end{equation}
Conversely, the presence of multiple MZMs complicates the hybridization dynamics as the evolution of the logical state is governed by the degree to which each pair of MZMs hybridize with each other throughout the process \cite{hodgeCharacterizingDynamicHybridization2025}. 
In this case, the system can then be characterized by the set of time dependent hybridization strengths $E_{ij}(t)$, 
as revealed by projecting the Hamiltonian onto the low-energy Majorana subspace:
\begin{equation}
    H_{\rm MZM}(t) = \frac{i}{4}\sum_{i,j}E_{ij}(t)\gamma_i(t)\gamma_j(t)\,.
\end{equation}

In general, to determine the exact final state produced by a braid, it is necessary to know the hybridization energies between all pairs of MZMs throughout the entire process, as their effects can compound in non-trivial ways\,\cite{hodgeCharacterizingDynamicHybridization2025}.
However, under the assumption that during each braid the only significant hybridization occurs between the two moving MZMs due to their proximity at key times, the braid between $\gamma_i$ and $\gamma_j$ will induce the additional dynamic component to the time evolution
\begin{equation}
    U_{\rm hyb}(T) = \exp\left(-\frac{\bar{E}_{ij}T}{2}\gamma_i\gamma_j\right)\,,
\end{equation}
where
\begin{equation}
    \bar{E}_{ij} = \frac{1}{T}\int_0^TdtE_{ij}(t)\,.
    \label{new_ebar}
\end{equation}

In the two MZM case, \eqref{new_ebar} simplifies to Eq.\,\EqEbar\ presented in the main paper by recognizing that as the only state in the low energy subspace $E_1(t)\equiv E_{12}(t)$. When discussing the hybridization energy during the braid $B_{ij}$, we are referring to the specific hybridization channel $E_{ij}$, as we assume all others are negligible, due to the larger separation between the static MZMs.
 
\section{Dynamic Ramping Protocol of Majorana Zero Modes for Braiding}

\subsection{$\sqrt{Z}$ and $\sqrt{X}$ Gates}

The ramping protocol we use to braid pairs of adjacent MZMs is adapted from Ref.\,\onlinecite{peetersEffectImpuritiesDisorder2024,mascotManyBodyMajoranaBraiding2023}. 
The full braid time, $T$, is split into six equal sections of period $T/6$, during each of which, a single section of wire has its chemical potential smoothly ramped from topological to trivial or vice versa according to Eq.\,\RampEq\ in the Methods section, moving the MZM along it. 
As briefly mentioned in the main paper, there are two variations of the braid, one for when the MZMs are separated by a topological region (e.g. $B_{12}$), and one for when they are separated by a trivial region (e.g. $B_{23}$), however, the MZMs move in the same general pattern for both processes.

\begin{figure}[h!]
    \centering
    \includegraphics[width=0.98\linewidth]{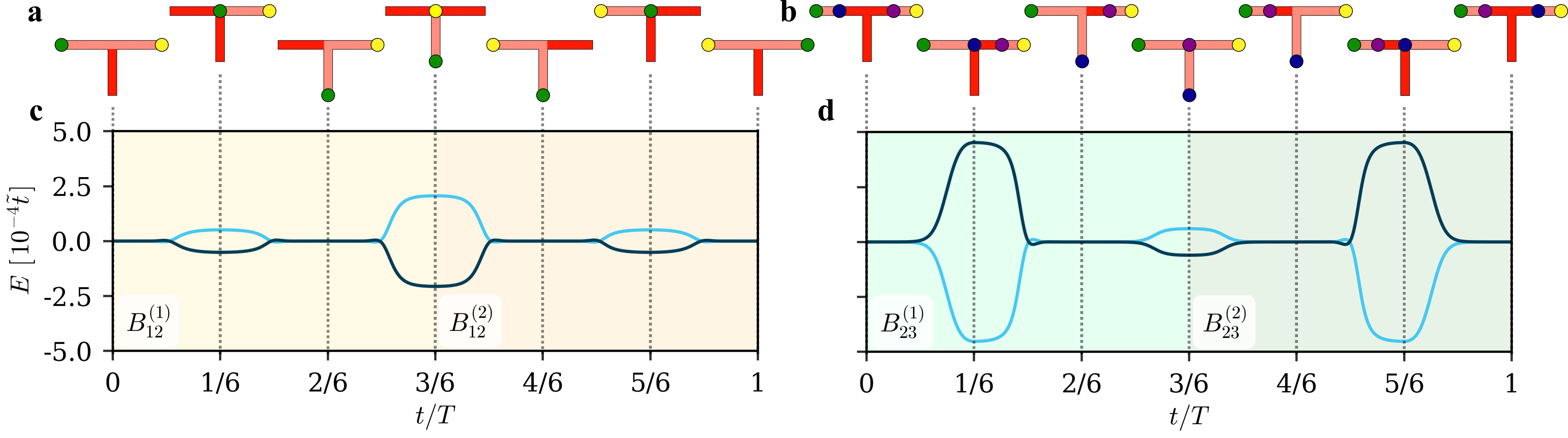}
    \caption{\textbf{Diagrams and hybridization energies for $\mathbf{B_{12}}$ and $\mathbf{B_{23}}$ ($\mathbf{\sqrt{Z}}$ and $\mathbf{\sqrt{X}}$ gates, respectively). a.} A schematic of the braid protocol for $B_{12}$. The light and dark red regions are where the system is topological and trivial, while the colored dots represent MZMs. \textbf{b.} The time-dependent hybridization energy of MZMs performing $B_{12}$. \textbf{c, d.} The same as \textbf{a} and \textbf{b} but for $B_{23}$.}
    \label{fig:Gate energies}
\end{figure}

Snapshots of the system at key times are shown in Fig.\,\ref{fig:Gate energies}\,\textbf{a}, where MZMs are represented by colored dots, and the light and dark red regions are topological and trivial, respectively. The green MZM at $t=0$ is advanced from its initial position to the center and then the bottom of the vertical leg. This allows the yellow MZM to move through the center and be placed in the original position of the green MZM, before the green itself is moved to the yellow MZM's initial position, completing the braid $B_{12}$. Fig.\,\ref{fig:Gate energies}\,\textbf{b} shows the similar braid, $B_{23}$, where the outer MZMs remain stationary as the inner ones perform the same sequence of movements.

In both braids, the hybridization energy across different times $t$, plotted in Fig.\,\ref{fig:Gate energies}\,\textbf{c} and \textbf{d}, is greatest when the MZMs are the closest: $t/T=1/6$, $3/6$, and $5/6$. The difference in magnitude between the peaks is a function of the system's leg lengths, given by $L_h$ and $L_v$, as well as the specific values of $\mu_{\rm topo}$ and $\mu_{\rm triv}$. The backgrounds are colored and labeled to highlight the half-braid components, $B^{(1)}_{ij}$ and $B^{(2)}_{ij}$, introduced in the main text. Due to the symmetry of the system across the vertical leg, both halves have the same total hybridization.

For more detail, the time-dependent chemical potentials each of these braids, as well as the associated hybridization energy and MZM density of states, performed on a 5T-junction can be seen in  Supplementary Movie 1 and Supplementary Movie 2.

\subsection{CNOT Gate}


To implement other Clifford gates, we perform the required braids sequentially, chaining the individual protocols defined above. We demonstrate this by performing a CNOT gate in Sec.\,\CnotSec, constructed by the sequence of braids $B_{12}B_{23}B_{12}B_{65}B_{34}B_{21}B_{12}B_{23}B_{12}$. 

It is important to note that this realization of the CNOT gate is restricted to when the parity of the physical Majorana bound states is even as it involves manipulating the ancilla qubit, whose physical occupation for a given logical state depends on the global parity. This can be specified via initialization\,\cite{Crawford2024Preparation,bordin2025}, and if the parity is odd, the CNOT gate is able to be enacted through the alternative process $B_{12}B_{23}B_{12}B_{65}B_\mathbf{43}B_{21}B_{12}B_{23}B_{12}$.


\section{Parity and the Pfaffian }

\begin{figure}[h!]
    \centering
    \includegraphics[width=0.98\linewidth]{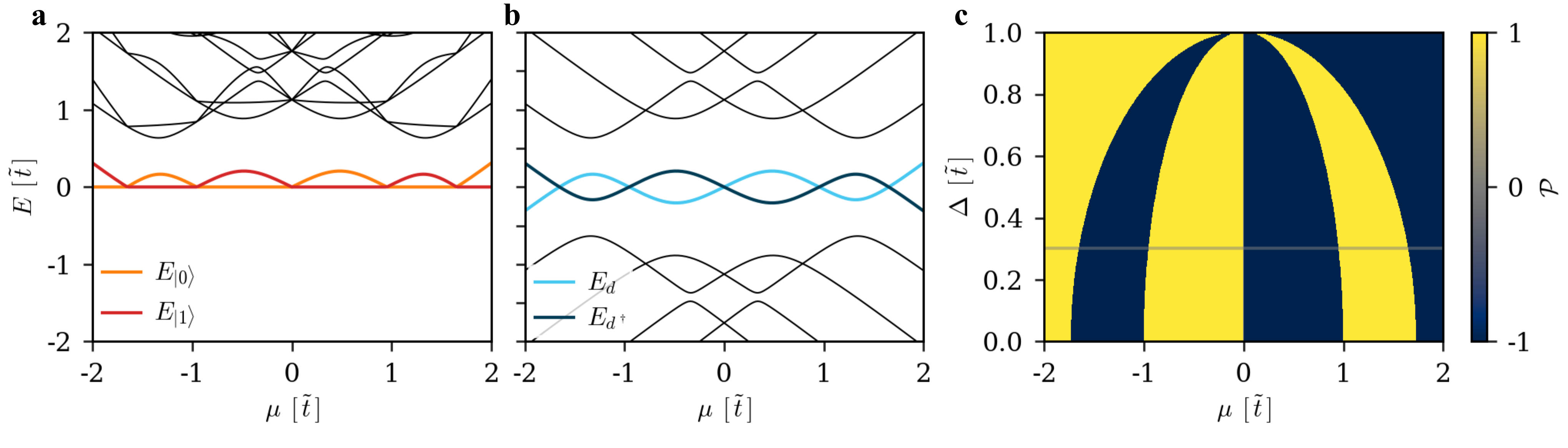}
    \caption{\textbf{Parity swaps on a Kitaev chain. a.} Many-body spectrum of a 5 site Kitaev chain for different values of $\mu$, shifted so that that the lowest energy is fixed at zero. The orange and red bands correspond to the states within the low energy subspace, while the black bands denote states where at least one bulk mode is occupied. \textbf{b.} Single-particle BdG spectrum for different values of $\mu$. The colored lines correspond to the energies required to fill or empty the Majorana bound state, while the black energies are for the bulk states. \textbf{c.} The parity of the ground state of the system for topological values of $\mu$ and $\Delta$. The gray line at $\Delta=0.3\tilde{t}$ corresponds to the superconducting order parameter used in panels in \textbf{a} and \textbf{b}.}
    \label{fig:MBS}
\end{figure}

An ideal Majorana bound state is a zero energy excitation, such that the unoccupied ground state and the occupied excited state are degenerate.
However, in most physical, i.e., finite systems, hybridization between the MZMs becomes unavoidable, breaking this degeneracy. Fig.\,\ref{fig:MBS} \textbf{a} plots the eigenvalues of the full many-body Hamiltonian for a five site Kitaev chain across a range of chemical potentials at $\Delta=0.3\tilde{t}$, shifted so that the lowest state is kept at $E=0$. The two lowest states (colored) correspond to the ground state and the state with the Majorana bound state excitation occupied. By taking the point $\mu=-2\tilde{t}$ we can label the lower energy $E_{|0\rangle}$ and the higher energy state $E_{|1\rangle}$. For the other values of $\mu$, we can connect the energies to these initial labels by considering their fermionic parity. Acting on the associated eigenstates with the parity operator,
\begin{equation}
    P = 
    \prod_j (1-2c_j^\dagger c_j^\pd),
\label{Eq: parity operator}
\end{equation}
we see that the two levels have opposite parities of $1$ and $-1$. This must be the case, as by definition $|1\rangle=d^\dagger|0\rangle$, implying that the two states differ in occupation number by one fermion. 

By imposing that the parity for each of the states is fixed for all values of $\mu$, we can use their parities to label them consistently. For some chemical potentials, the ordering of the states with respect to energy changes, such that $E_{|0\rangle}>E_{|1\rangle}$. This means that the state we labeled as $|0\rangle$ due to the choice of initial $\mu$ value, is no longer the ground state of the system, illustrating how the parity of the ground state is able to change throughout the course of the braid.

This can be compared to the BdG single-particle spectrum in Fig.\,\ref{fig:MBS}\,\textbf{b}, where the colored energies are associated with the single particle operators $d^\dagger$ and $d$. 
The energy of $d^\dagger$ corresponds to the energy difference between the many body states $|0\rangle$ and $|1\rangle$. As this difference in energy changes sign with the changes in ground state parity, $E_{d^\dagger}$ must also change sign as $\mu$ is increased.

Fig.\,\ref{fig:MBS}\,\textbf{c} shows the parity of the ground state of the same chain for various topological values of $\Delta$ and $\mu$, with the regions of $\mathcal{P}=1$ colored in yellow, and $\mathcal{P}=-1$ in blue. This structure reveals that the parity swaps occur for $\Delta<\tilde{t}$, and it has been shown previously \cite{leumerExactEigenvectorsEigenvalues2020,hegdeMajoranaWavefunctionOscillations2016} that the number and positions of the swaps depend on the length and parameters of the chain. However, for an isotropic chain with an odd number of sites, the parity always swaps at $\mu=0$, and is antisymmetric around the $\mu=0$ axis.

Although the many-body eigenstates have a well defined parity, the exponential scaling of the Hamiltonian with system size means that it is impractical to use it to calculate the sign of the hybridization. Instead we focus of on the single particle Hamiltonian which scales linearly.
 
 As shown by Kitaev\,\cite{kitaevUnpairedMajoranaFermions2001}, if the Hamiltonian is written in the form
\begin{equation}
    H=\frac{i}{4}\sum_{l,m}A_{lm}\gamma_l\gamma_m,
\label{eq: H mzm basis}
\end{equation}
the fermionic parity of the system is
\begin{equation}
    \mathcal{P} = \text{sgn}(\Pf(A)),
    \label{Eq: Pfaffian Parity}
\end{equation}
where $\Pf(A)$ is the Pfaffian of the skew-symmetric matrix $A$. 
The Pfaffian is the product of half the eigenvalues, one of each positive-negative pair produced by the BdG Hamiltonian. Therefore, its square is the determinant. 
For continuous changes in parameters, the bulk states cannot change sign without closing the gap, therefore any change in the sign of the total Pfaffian must be the result of the energy of a Majorana bound state changing sign --- indicating that the parity of the ground state has swapped.


To explore how the parity of a system's ground state can be manipulated to induce negative hybridization, we examine the Pfaffian of the single particle Kitaev chain Hamiltonian as given in Eq.\,\KitaevChainHamiltonian. 

In order to bring the Hamiltonian into the form of \eqref{eq: H mzm basis}, we represent it in the Majorana basis $H = \frac{i}{4}\Psi^\dagger A \Psi$ where the spinor $\Psi$ is given by $\Psi = \begin{bmatrix}
    \tilde{\gamma}_1&\tilde{\gamma}_2&\cdots&\tilde{\gamma}_{2N}
\end{bmatrix}^T$ using an alternative definition for the Majorana operators,
\begin{equation}
    \tilde{\gamma}_{2j-1} = c^\dagger_j+c_j\qquad {\rm and} \qquad\tilde{\gamma}_{2j} = i\left(c^\dagger_j-c_j\right).
\end{equation}

This ensures $A$ is a skew-symmetric, block-tridiagonal matrix given by
\begin{equation}
    A = \begin{bmatrix}
        B_1 & C_1 & & &\\
        -C^T_1 & B_2 & C_2 & &\\
         & -C^T_2 & \ddots & \ddots&\\
         & & \ddots & \ddots & C_{N-1}\\
         & &  & -C^T_{N-1} & B_{N}\\
    \end{bmatrix},
\end{equation}
where 
\begin{equation}
    B_j =\begin{bmatrix}
    0 & \mu_j\\
    -\mu_j & 0
\end{bmatrix}, \quad C_j = \begin{bmatrix}
    -{\rm Im}(\Delta_j) & \tilde{t}+{\rm Re}(\Delta_j)\\
    -\tilde{t}+{\rm Re}(\Delta_j) & {\rm Im}(\Delta_j)
\end{bmatrix}.
\end{equation}

The Pfaffian of a block matrix can be simplified into a product of smaller Pfaffians, as per the identity 
\begin{equation}
    \Pf\left(\begin{bmatrix}
    M & Q\\
    -Q^T & N
\end{bmatrix}\right) = \Pf(M)\Pf(N+Q^TM^{-1}Q),
\end{equation}
where $M$, $N$, and $Q$ are blocks which completely cover a larger skew-symmetric matrix.

If we take $M=B_1$ we see that 
\begin{equation}
    \Pf(A) = \Pf(B_1)\Pf\left(\begin{bmatrix}
        B_2 & C_2 & &\\
         -C^T_2 & \ddots & \ddots&\\
         & \ddots & \ddots & C_{N-1}\\
         &  & -C^T_{N-1} & B_{N}\\
    \end{bmatrix}+\begin{bmatrix}
        -C_1^T\\
        0\\
        \vdots\\
        0
    \end{bmatrix}B_1^{-1}
    \begin{bmatrix}
        C_1 & 0 &\dots &0
    \end{bmatrix}\right).
\end{equation}

As $B_1$ is a $2\times2$ matrix, by inspection $\Pf(B_1)=\mu_1$, leaving
\begin{equation}
    \Pf(A) = \mu_1\Pf\left(\begin{bmatrix}
        B'_2 & C_2 & &\\
         -C^T_2 & \ddots & \ddots&\\
         & \ddots & \ddots & C_{N-1}\\
         &  & -C^T_{N-1} & B_{N}\\
    \end{bmatrix}\right),
    \label{eq: reduced pfaffian}
\end{equation}
where $B'_2 = \begin{bmatrix}
    0 & \mu'_2\\
    -\mu'_2 & 0
\end{bmatrix}$ and $\mu'_2= \mu_2+\frac{\tilde{t}_1^2-|\Delta_1|^2}{\mu_1}$. The remaining Pfaffian in \eqref{eq: reduced pfaffian} is of exactly the same form as the original, allowing the procedure to be repeated iteratively. Finally, the full Pfaffian can be expressed as a product of the modified chemical potentials,
\begin{equation}
    \Pf(A) = \prod_{j=1}^N \mu_j',
\label{Eq: pfaffian product}\end{equation}
where $\mu_j' = \mu_j +\frac{\tilde{t}_{j-1}^2-|\Delta_{j-1}|^2}{\mu_{j-1}'}$ and $\mu_1' = \mu_1$. For a finite chain with constant parameters, our result agrees with previous findings\,\cite{leumerExactEigenvectorsEigenvalues2020,hegdeMajoranaWavefunctionOscillations2016}. 
However, in this form, it is easy to see that when $\tilde{t}^2=|\Delta|^2$, the sign of the parity is simply a product of the signs of the chemical potentials for each site.


The Hamiltonian for the T-junction is similar to that of a single chain, except for the fact that the first site of the vertical leg is connected to the center site, rather than the final horizontal site, which is labeled by $n$. 
In the transformed Hamiltonian, this moves the $C_n$ and $-C_n^T$ blocks off the tri-diagonal line, leading to a more complicated expression for the Pfaffian. 
However, by repeating a similar procedure, a closed form solution can be found. 
In the limit $|\tilde{t}|=|\Delta|$, this still converges to the product of the chemical potentials on each site.

These results justify how manipulating the sign of the chemical potential on even a small subset of the system during a braid can induce a parity swap, thus allowing for the presence of negative hybridization.

\section{Fidelity and Frobenius Norm}

\subsection{Fidelity}

To measure the accuracy of the simulated braids as quantum gates, we use the fidelity, as defined in Eq.\,\EqFidelity\ of the main text. 
For a given initial state, $|\Psi_{\rm initial}\rangle$, the corresponding target state, $|\psi_{\rm target}\rangle$, is defined so that $|\langle\Psi_{\rm target}|U_{\rm gate}|\Psi_{\rm initial}\rangle|^2=1$, where $U_{\rm gate}$ is the exact unitary intended to be performed.
By replacing $U_{\rm gate}$ with the numerically calculated time evolution of the braid, $U(T)$, we can see how close the generated unitary transformation is to the desired quantum gate. If $U(T)=U_{\rm gate}$, the fidelity will remain equal to one, however as they deviate, the fidelity will decrease, with a minimum of zero being reached when the final state is orthogonal to the target state.

As each gate transforms a given quantum state differently, their fidelities require different initial and target states so that the relevant information can be captured.
For example, a $Z$ gate acting on either the $|0\rangle$ or $|1\rangle$ states modifies their phase, which is undetectable when looking at the squared overlaps. Therefore, we use an initial state which is a linear combination of the two, $|\psi_{\rm{initial}}\rangle = \frac{1}{\sqrt{2}}\left(|0\rangle+|1\rangle\right)$, which maps to the different superposition $\frac{1}{\sqrt{2}}\left(|0\rangle-|1\rangle\right)$ when the $Z$ gate is applied to it.
    
The specific fidelities in the paper for $\sqrt{X}$ and $\sqrt{Z}$ gates are defined as 
 \begin{equation}
     \mathcal{F}_{\sqrt{X}}=|\langle -i|U(T)|0\rangle|^2\,; \quad
            \mathcal{F}_{\sqrt{Z}}=|\langle i|U(T)|+\rangle|^2,
 \end{equation}
where
$|+\rangle=\frac{1}{\sqrt{2}}\left(|0\rangle+|1\rangle\right)$, $|i\rangle=\frac{1}{\sqrt{2}}\left(|0\rangle+i|1\rangle\right)$, and $|-i\rangle=\frac{1}{\sqrt{2}}\left(|0\rangle-i|1\rangle\right)$.


In the static system, where no braids are being performed, each state should remain unchanged, implying that $U_{\rm gate}$ is equal to the identity matrix. Thus $|\psi_{\rm target}\rangle=|\psi_{\rm initial}\rangle$ for all initial states, giving a static fidelity of
\begin{equation}
    \mathcal{F}_{\rm static} = |\langle\psi_{\rm initial}|U(T)|\psi_{\rm initial}\rangle|^2.
\end{equation}

As each single-qubit gate corresponds to a uniform, global rotation of the Bloch-sphere, the choice of specific initial state is arbitrary (so long as it does not lie on the axis of rotation).
However, this is not the case for the CNOT gate, which applies a different rotation to the target qubit depending on the state of the control qubit. 
Therefore, a fidelity measurement with only a single initial state cannot fully represent the accuracy of the full unitary.
To circumvent this, in Fig.\,\FigFour\ we plot the fidelities of 1000 different, random initial states. Similarly to the single-qubit gates, each of the target states are defined by the initial state and the perfect CNOT gate, $|\Psi_{\rm target}\rangle = U_{\rm CNOT}|\Psi_{\rm inital}\rangle$, giving
\begin{equation}
    \mathcal{F}_{\rm CNOT} = \langle \Psi_{\rm inital}| U_{\rm CNOT}^\dagger U(T)|\Psi_{\rm inital}\rangle.
\end{equation}
Together, this ensemble demonstrates the expected fidelity of the CNOT gate, even as it depends on the specific initial state.

\subsection{Frobenius Norm}

An alternative to using the fidelity as defined above to measure the accuracy of the gates, is using the Frobenius norm. The Frobenius norm is able to  quantify the error in the produced time-evolution operator on a component-by-component basis, removing the need to choose a specific initial state.

The Frobenius norm of a matrix, $U$, is defined as
        \begin{equation}
            ||U||_F = \left[{\sum_{i,j}|U_{ij}|^2}\right]^{1/2},
        \end{equation}
where the lowest value it can take is 0, when all of the components are zero.

The error of the CNOT gate, $\mathcal{E}$, can be defined using the Frobenius norm as
        \begin{equation}
            \mathcal{E}_{\rm CNOT} = ||U(T) - U_{\rm CNOT}||_F,
        \end{equation} 
where $U(T)$ is the time-evolution operator of the simulated gate and $U_{\rm CNOT}$ is the ideal CNOT gate operator. When $U(T) = U_{\rm CNOT}$ the simulated gate is perfectly accurate, giving $\mathcal{F_{\rm CNOT}}=1$ and $\mathcal{E}_{\rm CNOT} = 0$.

Unfortunately, the Frobenius norm is unbounded from above, so unlike the original fidelity where a value of 0 corresponds to a final state which is orthogonal to the target, the specific value the norm takes does not directly correspond to an observable property of the final state. However, it is still possible to compare between the norms of different operations as $\mathcal{E}_{\rm CNOT}$ is a monotonically increasing function of the inaccuracy, so a smaller norm will always correspond to a more accurate gate.

We can also consider the error associated with the static system, given by
\begin{equation}
    \mathcal{E}_{\rm static} = ||U(T) - \mathbb{1}||_F,
\end{equation}
where $\mathbb{1}$ is the identity gate and the specific time-evolution operator assumes the system remains in its initial configuration.

Additionally, we can isolate the component of the error corresponding to the loss of quantum information to excited states, $\mathcal{L}$. We use the Frobenius norm to test the unitarity of the time-evolution operator, quantified by the following:
\begin{equation}
    \mathcal{L} = \frac{1}{\sqrt{\mathcal{N}}}||\mathbb{1}-U(T)^\dagger U(T)||_F\,,
\end{equation}
where $\mathcal{N}$ is a normalization factor equal to the dimension of the logical subspace which the braiding operators act on. For example, the CNOT with two qubits has $\mathcal{N}=2^2=4$.
As the loss is dominated by diabatic effects, the remaining error is predominantly due to hybridization. Therefore, by comparing $\mathcal{E}$ and $\mathcal{L}$ for the corrected gates, we are able to determine the effectiveness of the protocol.

\begin{figure}[h!]
            \centering
            \includegraphics[width=0.98\linewidth]{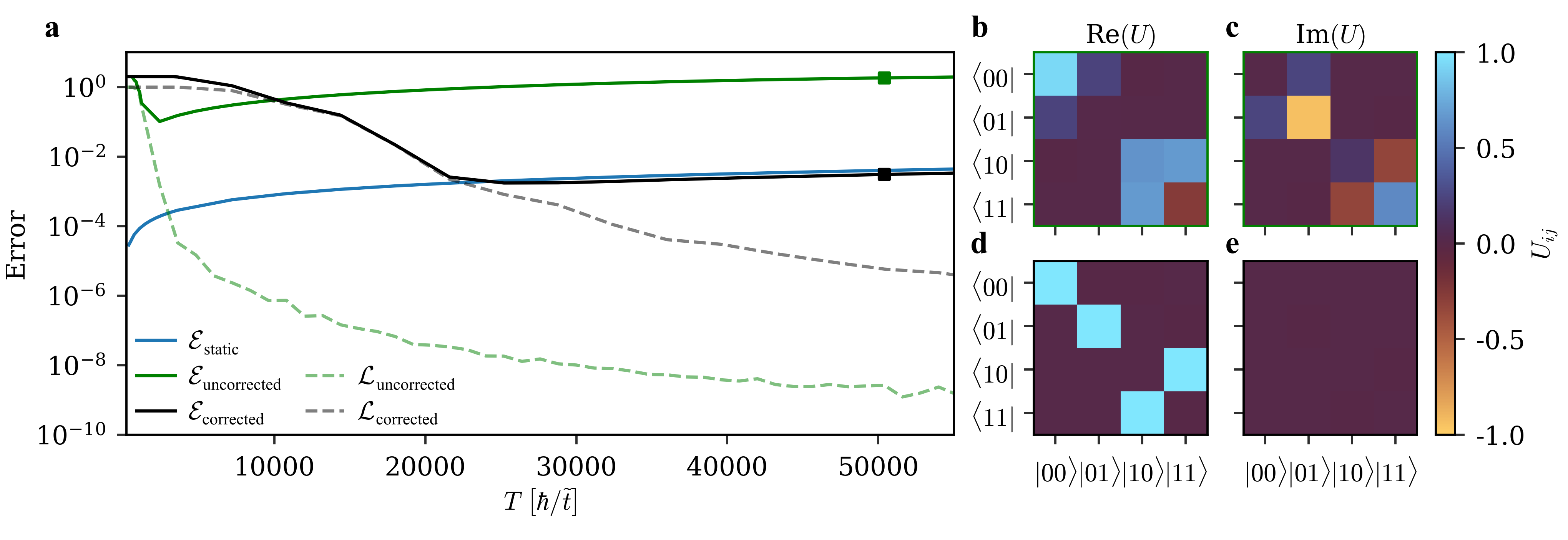}
            \caption{\textbf{CNOT gate error measured using the Frobenius norm. a.} The error, $\mathcal{E}_{\rm CNOT}$, of the time evolution of the uncorrected (green) and corrected (black) CNOT gates as a function of braid time, $T$. The corresponding diabatic losses, $\mathcal{L}$, are plotted as dashed lines, providing a lower bound for the gate errors. Additionally, $\mathcal{E}_{\rm static}$ is plotted for the static case which shows how in the adiabatic limit, the accuracy of the corrected CNOT gate depends on the system's innate hybridization. \textbf{b, c.} The real and imaginary components of the uncorrected CNOT gate at $T\approx50000\hbar/\tilde{t}$, indicated by the green square in \textbf{a}. 
            \textbf{d, e.} The same as \textbf{b, c} but for the corrected gate, marked by the black square in \textbf{a}.
            }
            \label{fig:CNOT Frobenius}
        \end{figure}

These errors, as a function of $T$, are plotted for the corrected (black) and uncorrected (green) CNOT gates in Fig.\,\ref{fig:CNOT Frobenius}\,\textbf{a}, where they qualitatively match the results in Fig.\,\FigFour. The loss, $\mathcal{L}$, of each gate is plotted as a dashed line of the same color, revealing the constraint imposed by the lower speed limit. The uncorrected results quickly depart from the diabatic limit as $T$ increases and the effects of hybridization become relevant. Meanwhile, the corrected results closely follow the diabatic loss until the error due to hybridization in the static system, shown in blue, surpasses it.

To provide intuition into how specific values of $\mathcal{E}_{\rm CNOT}$ relate to the physical gate, the unitaries for the corrected and uncorrected gates at $T\approx50000\hbar/\tilde{t}$, marked by the black and green squares respectively, are shown in Fig.\,\ref{fig:CNOT Frobenius}\,\textbf{b-d}. The uncorrected gate is shown in \textbf{b} and \textbf{c}, with the real components on the left and the imaginary on the right, while \textbf{d} and \textbf{e} show the same for the corrected gate. The uncorrected gate with $\mathcal{E}_{\rm CNOT}\approx10^0$ appears to be almost random, while the corrected gate with $\mathcal{E}_{\rm CNOT}\approx10^{-3}$ conforms well to the expected CNOT gate.


\section{Extensions to Negative Hybridization Enforced by a Local Gate}






\subsection{Effect of the Magnitude of the Superconducting Gap}
\begin{figure}[!h]
    \centering
    \includegraphics[width=0.98\linewidth]{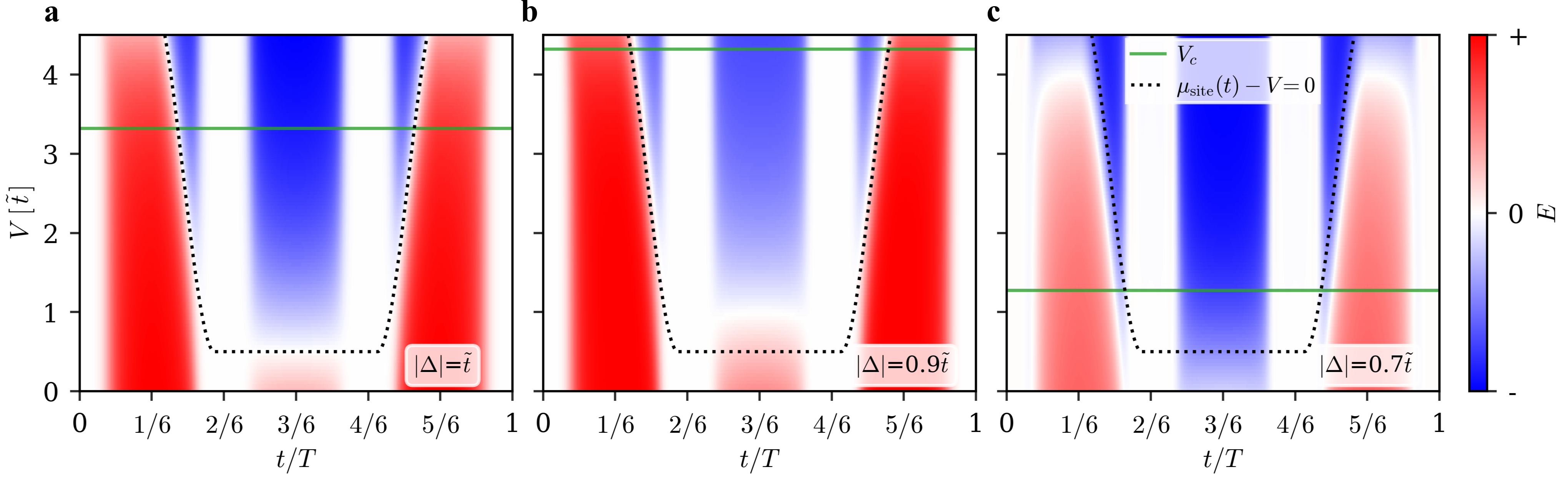}
    \caption{\textbf{Parity swaps induced by local gating voltage, away from the $\mathbf{|\Delta|=\tilde{t}}$ limit.} \textbf{a.} Hybridization energy throughout a $\sqrt{Z}$ gate with local gating applied to the site below the center with strength $V$, for $|\Delta|=\tilde{t}$. The green line marks $V_c$, the critical applied voltage where $\bar{E}=0$. The dotted line signifies where $\mu_{\rm site}(t)-V=0$. \textbf{b,} \textbf{c.} The same as \textbf{a}, but for $|\Delta|=0.9\tilde{t}$ and $|\Delta|=0.7\tilde{t}$, respectively. \textbf{c} shows the same data as Fig.\,\FigTwo\,\textbf{a}.}
    \label{fig:site23 different delta}
\end{figure}
As shown in \eqref{Eq: pfaffian product}, when the magnitude of the superconducting order parameter is equal to that of the hopping, changing the sign of one site's chemical potential is sufficient to induce negative hybridization. However, as the magnitude of $\Delta$ is reduced, the parity of the ground state is no longer trivially ascertainable, as it then depends on the exact geometry and chemical potential profile of the system.

Examining the hybridization energy for a $\sqrt{Z}$ gate where $|\Delta|=\tilde{t}$ in Fig.\,\ref{fig:site23 different delta}\textbf{a}, the dotted line indicating when the chemical potential on the site with the applied voltage, given by $\mu_{\rm site}(t)-V$, perfectly coincides along the boundary between the positive and negative energies. This agrees with \eqref{Eq: pfaffian product} as it is precisely when the parity swap is predicted to occur, however, as $\Delta$ decreases, the induced parity swaps deviate from $\mu_{\rm site}(t)-V$.
Fig.\,\ref{fig:site23 different delta}\textbf{b} and \textbf{c} show the same gate with $|\Delta|=0.9\tilde{t}$ and $|\Delta|=0.7\tilde{t}$, respectively, and demonstrate initial parity swap occurring before and after the calculated time.

This leads to very different values for $V_c$, the critical voltage marked by the green lines where $\bar{E}=0$, highlighting the difficulty in attempting to use this method to predict the needed voltage to be applied to correct a given braid. However, the distinctive oscillation in fidelity as a function of $V$ and $T$, shown in Fig.\,\FigTwo\,\textbf{c}, can be used to tune the voltage to $V_c$ for a given system, assuming the presence of the parity swaps is similar enough to those predicted in the $|\Delta|=\tilde{t}$ limit.

\subsection{Uniform Extended Gating Voltage}

\begin{figure}[h!]
        \centering
        \includegraphics[width=0.98\linewidth]{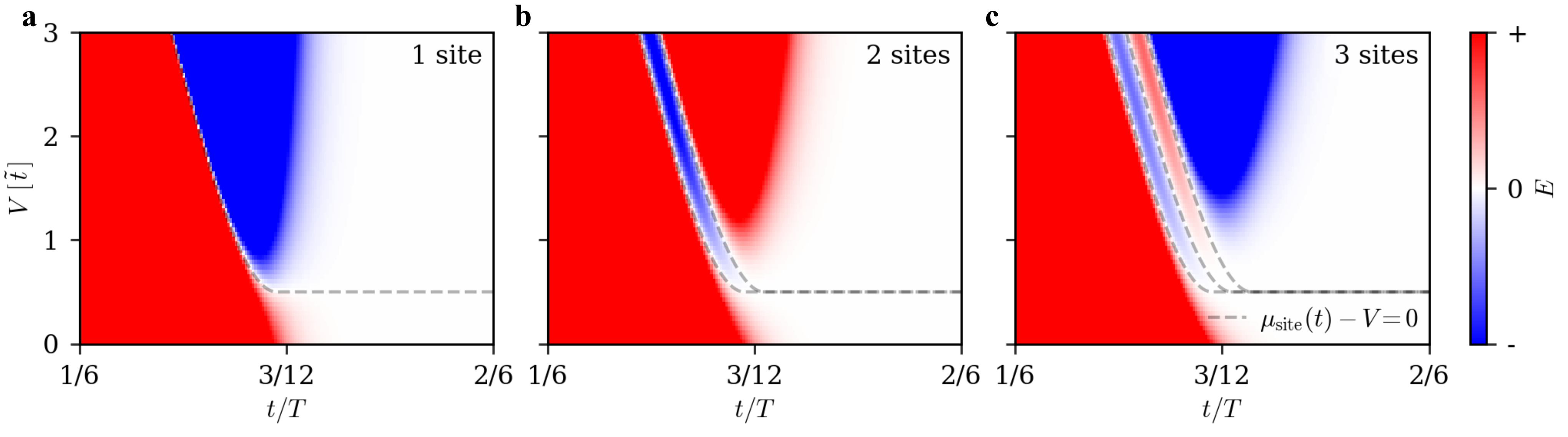}
        \caption{\textbf{Parity swaps for gating voltages extended over multiple sites. a.} The sign of hybridization energy for various times around the initial parity swap during a $\sqrt{Z}$ gate where $|\Delta|=\tilde{t}$ and a gating voltage with strength $V$ is applied to the site below the center. The points where $\mu_{\rm site}(t)-V=0$ are marked by a dotted line. \textbf{b, c.} The same as \textbf{a} with the voltage also being applied to one and two more sites below the original, respectively.}
        \label{fig:site_23_multi}
    \end{figure}

As depending on the system, applying the gating voltage to a single site may be difficult, it is informative to consider the effects of extending it over multiple sites. The simplest case is for a gating voltage of uniform strength being applied to the adjacent sites down from the initially considered site. Fig.\,\ref{fig:site_23_multi} shows the sign of the hybridization energy for the times around the initial parity swap of a $\sqrt{Z}$ gate with $|\Delta|=\tilde{t}$, for voltages of strength $V$ applied to one, two, and three sites.

As the chemical potential of each site changes sign one after the other as the braiding protocol sequentially lowers the chemical potential down the vertical leg, the parity of the system flips repeatedly as seen in Fig.\,\ref{fig:site_23_multi}\,\textbf{b} and \textbf{c}. Thus, the number of sites the potential extends over plays a significant role in determining the value of $\bar{E}$. For a low delay coefficient, $\alpha$, the parity of the system will rapidly alternate causing the impact on $\bar{E}$ to be minimal. As the largest contributions to the hybridization come before any of the sites change sign, and after they have all changed sign, it is necessary for the parity at the end of the ramp to be opposite that of the start to achieve significant negative hybridization.
For potentials which extend over an odd number of sites as in Fig.\,\ref{fig:site_23_multi}\,\textbf{c}, the hybridization is qualitatively similar to that of a single site. However for an even number of sites, shown in Fig.\,\ref{fig:site_23_multi}\,\textbf{b}, the successive parity swaps leave the system in the same parity as it was initially, preventing the required negative hybridization to find $\bar{E}=0$.

\subsection{Gaussian Extended Gating Voltage}

Instead of considering a uniform gate potential extended over several sites, we focus now on the somewhat more experimentally realistic situation
where the gating voltage is applied to a targeted point with strength $V$, but leaks to other sites, leading to a voltage which decreases with distance. For simplicity, we model this as a Gaussian distribution, where the strength on site $j$ is given by
\begin{equation}
    V_j = Ve^{-d_j^2/2\sigma^2},
\end{equation}
where $V$ is the strength of the potential at the target location, $\sigma$ characterizes the localization of the potential, and $d_j$ is the rectilinear distance between targeted position and the $j^{\rm th}$ site.

\begin{figure}[h!]
    \centering
    \includegraphics[width=0.98\linewidth]{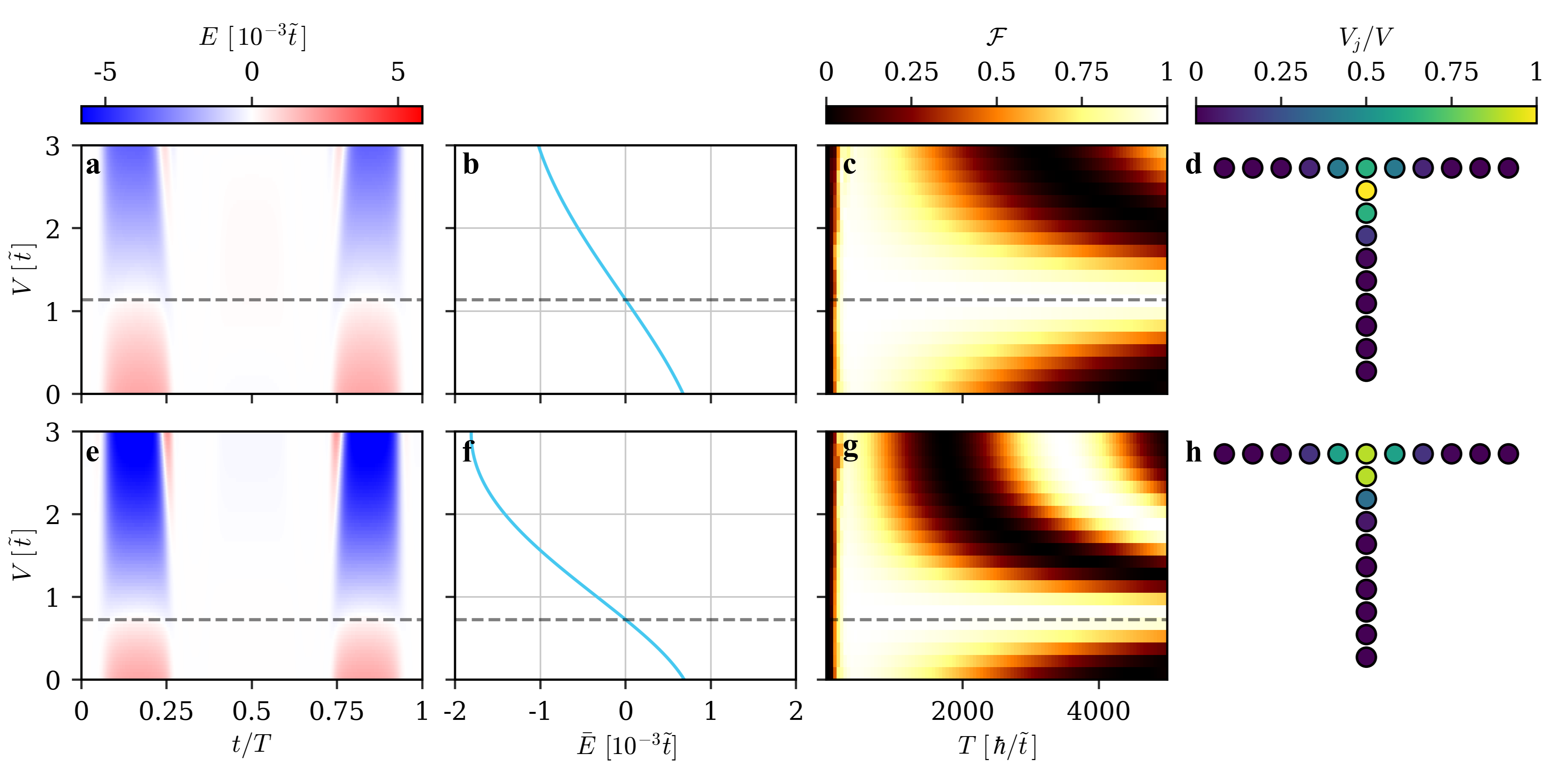}
    \caption{\textbf{Negative hybridization induced by a Gaussian potential. a.} The sign of the hybridization energy induced throughout a $\sqrt{Z}$ gate where a Gaussian potential of $\sigma=1.05$ is applied at various strengths, $V$, centered on the site at the top of the vertical leg. \textbf{b.} The average hybridization energy of the braid, $\bar{E}$, as a function of $V$. The critical voltage, $V_c$, where $\bar{E}=0$ is indicated by the dashed line across \textbf{a},\textbf{b} and \textbf{c}. \textbf{c.} The fidelity, $\mathcal{F}$, of the gate as a function of $T$ and $V$. For voltages around $V_c$, $\mathcal{F}$ remains close to 1 for large $T$. \textbf{d.} An illustration of the spatial profile of the Gaussian potential, where the $j^{th}$ site is colored corresponding to the the potential applied to it, $V_j$, as a fraction of $V$. \textbf{e-h.} The same as \textbf{a-d}, but the Gaussian potential is centered in between the top site of the vertical leg and the center site.}
    \label{fig:gaussian}
\end{figure}

We find that similar to the single-site potential, the voltage of the Gaussian profile can be tuned to obtain $\bar{E}=0$ for a $\sqrt{Z}$ gate, over a range of $\sigma$ values. In Fig.\,\ref{fig:gaussian}\,\textbf{a-c} we plot the hybridization energy, $\bar{E}$, and fidelity as a function of voltage strength, similar to Fig.\,\FigTwo, where the gating potential takes Gaussian profile centered on the top site of the vertical leg with $\sigma=1.05$, as shown in Fig.\,\ref{fig:gaussian}\textbf{d}. Negative hybridization is still induced in wedged regions that increase with $V$, however the transition from positive to negative $\bar{E}$ is dominated by a new lobe-like feature. Due to its presence, $V_c\approx1.1\tilde{t}$, allowing for the braid fidelity to remain close to 1 for all sufficiently adiabatic $T$.

We also investigate the effects of centering the potential in between the top site on the vertical leg, and the center site. For low $\sigma$ values, this is similar to a uniform two site potential, which as discussed above, is unable to produce a sufficient amount of negative hybridization to reach $\bar{E}=0$. However, for larger values, including $\sigma=1.05$ as shown in in Fig.\,\ref{fig:gaussian}\textbf{e-h}, the hybridization during the braid is comparable to when the potential is centered directly on the site. This allows for an even lower $V_c$, approximately $0.7\tilde{t}$.

These are both promising results, demonstrating that the applied potential does not have to be perfectly confined to a single site to produce the negative hybridization required to drastically reduce $\bar{E}$.


\section{Details on the Symmetric Braids}
\subsection{Symmetry Considerations}

 The parity of the ground state of a Kitaev chain with respect to $\mu$ and $\Delta$ is symmetric around $\mu=0$ for an even number of sites and antisymmetric for an odd number of sites, as seen in Fig.\,\ref{fig:MBS}\,\textbf{c}. See Section \ref{sec: spinful systems} for a discussion of other topological superconductor models. Due to the parity swaps inducing negative hybridization, a process which linearly changes the chemical potential of a chain with an odd number of sites from $\mu$ to $-\mu$ should have $\bar{E}=0$.

The single particle Hamiltonian for a Kitaev chain as given in Eq.\,\KitaevChainHamiltonian\ can be written to emphasize its dependence on the parameters $\tilde{t}$, $\mu$, and $\Delta$ as follows:
\begin{equation}
    H_{\rm K}(\tilde{t},\mu,\Delta) = \sum_{j=0}^N -\mu_j(t)c_j^\dagger c_j^\pd + \sum_{j=0}^{N-1} -\tilde{t}\,c_{j+1}^\dagger c_j^\pd + \Delta c^\dagger_{j+1} c_j^\dagger +{\rm H.c.}\,.
\end{equation}
Then, when acted upon by the particle-hole transformation
\begin{equation}
    c_j^\pd \rightarrow (-1)^j c_j^\dagger \qquad {\rm and} \qquad c_j^\dagger \rightarrow (-1)^j c_j^\pd\,,
    \label{eq: transformation}
\end{equation}
the Hamiltonian becomes
\begin{equation}
    H_{\rm K}(\tilde{t},\mu,\Delta) \rightarrow -\mu N +\sum_{j=0}^N \mu_j(t)c_j^\dagger c_j^\pd + \sum_{j=0}^{N-1} -\tilde{t}\,c_{j+1}^\dagger c_j^\pd + \Delta^* c^\dagger_{j+1} c_j^\dagger +{\rm H.c.} = H_{\rm K}(\tilde{t},-\mu,\Delta^*)-\mu N\,.
\end{equation}

Disregarding the constant shift $-\mu N$ which has no effect on the physics, we can see that a chain with chemical potential $\mu$ and superconducting order parameter $\Delta$ is related by a unitary transformation to the chain with parameters $-\mu$ and $\Delta^*$. For a chain with a constant superconducting order parameter, the phase can be gauged away, and thus does not effect the spectrum. On a T-junction where the horizontal and vertical legs have different phases, the global swap between $\Delta$ and $\Delta^*$ can be thought of as a transition from an underlying $p_x+ip_y$ to a $p_x-ip_y$ superconductor, which have the same spectra. The bulk states only depend on $|\Delta|$, while the energies of the subgap states located at the junctions depend on the difference in phase between adjacent sites\,\cite{aliceaNonAbelianStatisticsTopological2011} which remain the same under the change. Thus, for the systems we consider, the energy levels are conserved under a swap from $\mu$ to $-\mu$.


However, if we examine how the parity operator, as defined in \eqref{Eq: parity operator}, acts under the transformation \eqref{eq: transformation} we see that
\begin{equation}
    P \rightarrow \prod_j(-1+2c_j^\dagger c_j) = -P.
    \label{eq: parity swap}
\end{equation}
Together, the invariance of the BdG spectrum and the change in parity demonstrate that the hybridization of the two systems of opposite chemical potentials are equal in magnitude, but opposite in sign, justifying the occurrence of negative hybridization.

For the swap in chemical potential to induce negative hybridization, the affected region needs to have the correct number of sites, otherwise the procedure will lead to a doubling of the total hybridization, rather than its elimination. 
For the $\sqrt{Z}$ gates which only involve one pair of MZMs, the total number of swapped sites must be odd as discussed above. However, for gates involving two different pairs of MZMs such as the $\sqrt{X}$, the number of sites needed is doubled, making it even.

As each junction is symmetric across the vertical legs, the number of horizontal sites, including the center one, must be odd, thus constraining the parity of the vertical legs. The vertical legs involved in $\sqrt{Z}$ gates therefore have an even number of sites, while the ones involved in $\sqrt{X}$ gates have an odd number; both of which are satisfied by choosing an even value for $L_v$.
Ensuring that the system is comprised of the correct number of sites is trivial for certain physical realizations of the Kitaev chain such as MSH systems\,\cite{kim2018,schneider21} and tuned quantum dots\,\cite{leijnseParityQubitsPoor2012,Liu2022}. However, it is currently unclear how exactly arguments about site numbers transfer to the continuum limit relevant for devices such as superconductor-semiconductor nanowires\,\cite{mourikSignaturesMajoranaFermions2012} which is beyond the scope of this paper.


As the process of changing the sign of the chemical potential affects the MZMs, it is potentially a source of error within the whole gate. 
Therefore, it is useful to minimize the use of this operation where possible.
In the systems we consider, where the MZMs being braided hybridize exponentially more than any others, the fidelity of the gates can still be drastically improved, even when neglecting to correct for the error introduced by the stationary MZMs as shown in Fig.\,\FigFour\ and Fig.\,\ref{fig:CNOT Frobenius}.
Thus, our proposed protocol for the corrected gates only change the sign of the chemical potential of the sites which are directly involved in a given braid as demonstrated in Supplementary Movies 3 and 4.

This is sufficient to induce negative hybridization, which can be seen by considering the limit where the hybridization between the active and spectating MZMs goes to zero. In that case, the Hamiltonian can be split into two independent regions while preserving the low energy dynamics. Changing the chemical potentials of the portion containing the active MZMs is therefore effectively the same as changing the sign of $\mu$ for a whole system, inducing the required swap in parity as shown by \eqref{eq: parity swap}.
As the MZMs begin to hybridize more strongly and this approximation breaks down, we depart from the original assumption that the spectating MZMs are well defined on the time scale of a braid, leading to the qubits decohering even before the application of any gates.

However, it is important to note that for systems with $\mu_{\rm triv}$ somewhat close to the critical value of $2\tilde{t}$, the MZM wavefunctions will penetrate into neighboring trivial regions, making the hybridization energy depend on the chemical potentials of those sites. To account for this, our corrected $\sqrt{Z}$ protocol includes changing $\mu_{\rm triv}$ to $-\mu_{\rm triv}$ for the adjacent sites, ramping from the topological region so as to avoid the creation of new MZMs. This also inspires the use of `buffer' sites at each end of the horizontal backbone of the larger systems, to ensure the MZMs are always adjacent to trivial sites, rather than being at the end of a chain, so that the hybridization through the left and right legs remains identical. As the MZMs are exponentially localized, it is sufficient to consider only a small number of buffer sites; the simulated CNOT shown in Fig.\,\FigFour\ and Fig.\,\ref{fig:CNOT Frobenius} only uses four buffer sites.

Conversely, in the locations where the MZMs do not hybridize, the chemical potential can be left unchanged with minimal effect on the energy levels. This includes the vertical legs which lie outside the path the MZMs take. In a $\sqrt{X}$ gate, only the middle vertical leg needs to change sign, as the MZMs hybridize while they have significant wavefunction overlap on them, whereas the other two legs, which stay trivial, can remain positive as shown in Supplementary Movies 3 and 4.

\subsection{Negative Hybridization Routine}

 The modified braiding protocols are comprised of three sections: the original braid which accumulates hybridization $\bar{E}_+$, a trivial `braid' with negative chemical potentials which has hybridization $\bar{E}_-$, and the transitions between the system having positive and negative chemical potentials with hybridization $\bar{E}_{\rm swap}$. For the full protocol to have no net hybridization, we require $\bar{E}_+=-\bar{E}_-$ and $\bar{E}_{\rm swap}=0$.

 \begin{figure}[h!]
    \centering
    \includegraphics[width=0.98\linewidth]{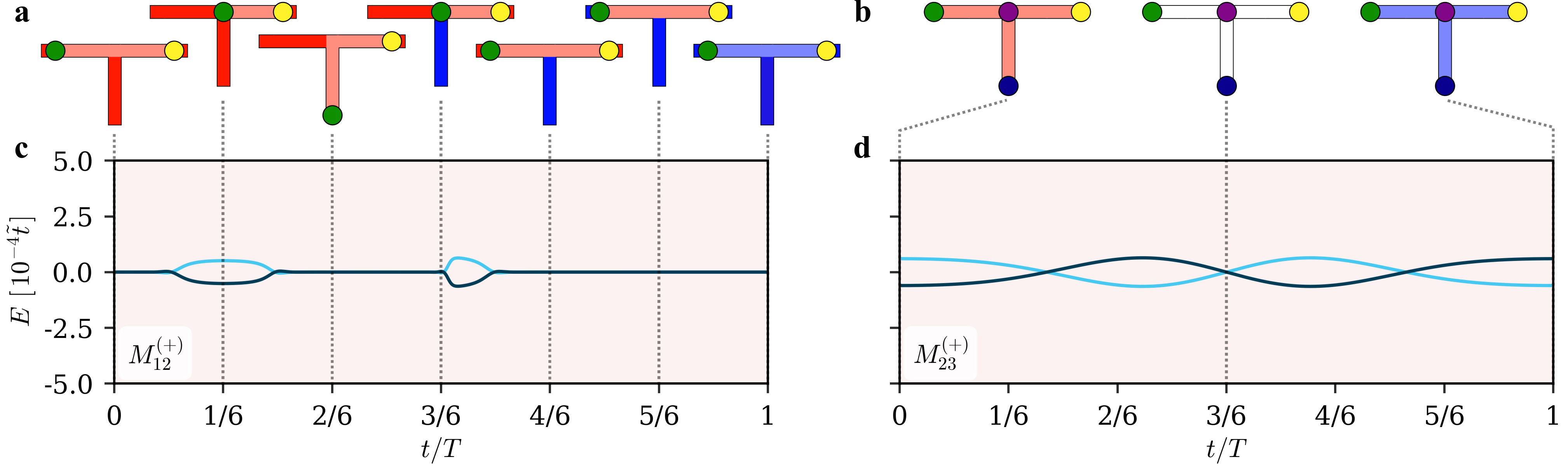}
    \caption{\textbf{Diagrams and hybridization energies of $M^{(+)}_{12}$ and $M^{(+)}_{23}$, the processes which transition the system from positive to negative chemical potentials. a.} Diagrammatic representations of the system at key times as it undergoes $M^{(+)}_{12}$. The lighter regions are topological while the darker regions are trivial, with red (blue) denoting a  positive (negative) chemical potential. The colored dots represent the MZMs pinned to the boundaries between regions. \textbf{b.} Diagrammatic representations of the system at key times as it undergoes $M^{(+)}_{23}$. Initially, all sites are at $\mu_{\rm topo}$, indicated by the light red color, before eventually becoming $-\mu_{\rm topo}$ shown by the blue in the final diagram. Halfway through the process, the chemical potentials are zero, which is shown by the white regions in the middle diagram. \textbf{c.} The hybridization energy during an example $M^{(+)}_{12}$ operation. \textbf{d.} The hybridization energy during an example $M^{(+)}_{23}$ operation. Due to the symmetry of the process, the hybridization of the first and second halves cancel.}
    \label{fig:swapping procedures}
\end{figure}

The original braids shown in Fig.\,\ref{fig:Gate energies} and Supplementary Movies 1 and 2. The trivial operations are constructed from the halves of the unmodified braids and are defined as $B_{ij}^{(1)} {B_{ij}^{(1)}}^{-1}$ or ${B_{ij}^{(2)}} {B_{ij}^{(2)}}^{-1}$.

The swapping procedures, $M_{ij}^{(\pm)}$, differ for the braids between even and odd numbered junctions due to the alternating configuration of topological and trivial regions. 
Fig.\,\ref{fig:swapping procedures}\,\textbf{a} sketches the system at key times during $M^{(+)}_{12}$ used in the $\sqrt{Z}$ gate. The lighter regions are topological while the darker regions are trivial, with red (blue) denoting a  positive (negative) chemical potential. In order to avoid the creation of new MZMs while tuning $\mu$ into the topological regime, which would potentially lead to a loss of quantum information, the swapping procedures are designed to only extend and retract the initial topological regions without generating any new ones. Starting from the initial configuration, one MZM is moved down the vertical leg so that on its return, the sites can be ramped to $-\mu_{\rm triv}$. Then, the MZMs extend into buffer regions allowing them to swap to $-\mu_{\rm triv}$, and finally the remaining topological sites are linearly ramped to $-\mu_{\rm topo}$.

Fig.\,\ref{fig:swapping procedures}\,\textbf{b} shows the same for $M^{(+)}_{23}$, used in the $\sqrt{X}$ gates. Starting the system in the middle of $B_{23}$, all of the sites that need to be swapped are topological, meaning a global ramp from $\mu_{\rm topo}$ to $-\mu_{\rm topo}$ is able to change the sign of all of the chemical potentials.

Looking at the associated time-dependent spectra in Fig.\,\ref{fig:swapping procedures}\,\textbf{c} and \textbf{d}, it can be seen that while $M^{(+)}_{23}$ has no net hybridization, due to the antisymmetric nature of the process around $t/T=1/2$, $M^{(+)}_{12}$ has non-zero hybridization, given by $\bar{E}_{\rm swap}^+$. 
To see why this is feasible, we must consider the inverse processes $M^{(-)}_{23}$ and $M^{(-)}_{12}$, which return the negative chemical potentials back to their initial values. 
Due to its simple nature $M^{(-)}_{23}$ is equivalent to the inverse of ${M^{(+)}_{23}}$. 
However, this is not the case for $M^{(-)}_{12}$, which is instead defined by performing the same protocol as $M^{(+)}_{12}$ with $-\mu$ instead of $\mu$. 
This complementary process has energy $E_{\rm swap}^-=-E_{\rm swap}^+$ due to the parity swap, meaning that together they contribute no net hybridization to the braid as required.

\begin{figure}[h!]
    \centering
    \includegraphics[width=0.98\linewidth]{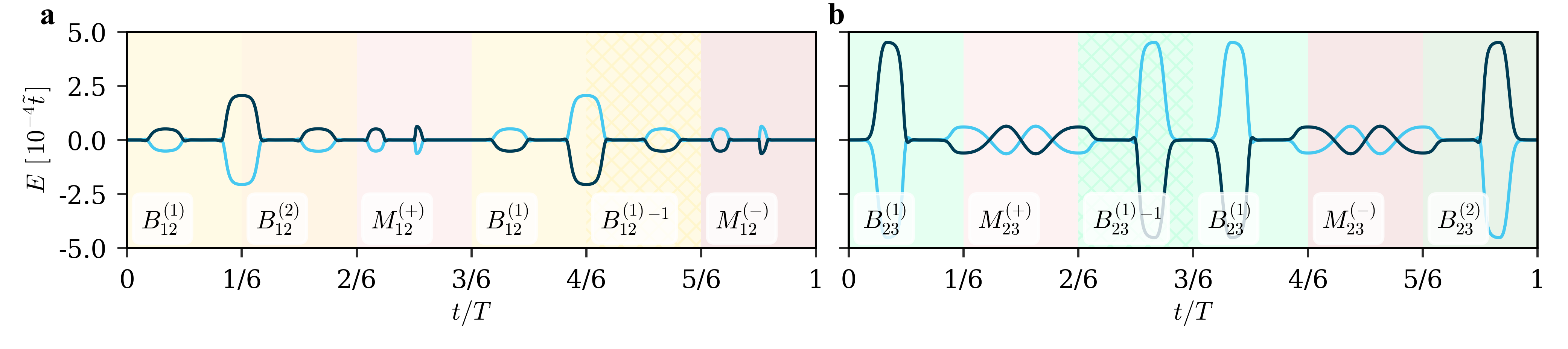}
    \caption{\textbf{Hybridization of the full corrected braids. a.} The time-dependent hybridization energy for the corrected $\sqrt{Z}$ gate. Each component is labeled with appropriately and distinguished by a unique color. \textbf{b.} Same as \textbf{a} but for the corrected $\sqrt{X}$ gate.}
    \label{fig:full corrected braids}
\end{figure}

The hybridization throughout the full $\sqrt{X}$ and $\sqrt{Z}$ gates are shown in Fig.\,\ref{fig:full corrected braids}. 
For both braids, the negative trivial braids are inserted between the chemical potential swaps. For the $\sqrt{Z}$ gate, the negative braid is performed after the original , while for the $\sqrt{X}$ gate, the negative braid occurs in the middle.

The corrected braids can be defined as 
\begin{equation}
    \tilde{B}_{12} = M^{(-)}_{12} {B^{(1)}_{12}}^{-1} B^{(1)}_{12} M^{(+)}_{12} B^{(2)}_{12} B^{(1)}_{12}\,,\quad \tilde{B}_{23} =B^{(2)}_{23} M^{(-)}_{23} B^{(1)}_{23} {B^{(1)}_{23}}^{-1} M^{(+)}_{23} B^{(1)}_{23}\,,
\end{equation}
where each component is colored and labeled in Fig.\,\ref{fig:full corrected braids}. Animated examples of the $\sqrt{X}$ and $\sqrt{Z}$ gates are shown in Supplementary Movies 3 and 4.

\section{Negative Hybridization in slightly Asymmetric Braids \label{Effects of introducing asymmetry}}

We note that due to the proposed protocol relying on the positive and negative braids accruing equal but opposite amounts of hybridization, unlike in the thermodynamic limit where there is no hybridization for the entire duration, the outcome is path dependent. While we are able to demonstrate the effectiveness of such a protocol under the assumption that all parameters can be perfectly controlled, as shown in the main text, it is important to consider how robust the procedure is in the face of the inevitable miscalibrations present in any experimental realization. In Fig.\,\FigThree\ we investigated the robustness against errors in the routine by analyzing the results of corrected $\sqrt{X}$ and $\sqrt{Z}$ gates when the topological chemical potentials used in the negative half, $\mu^-_{\rm triv/topo}$, are not exactly the same magnitude as the positive ones. The induced asymmetry led to oscillations in the fidelity as $\bar{E}$ was no longer exactly at zero.

\begin{figure}[!h]
        \centering
        \includegraphics[width=0.98\linewidth]{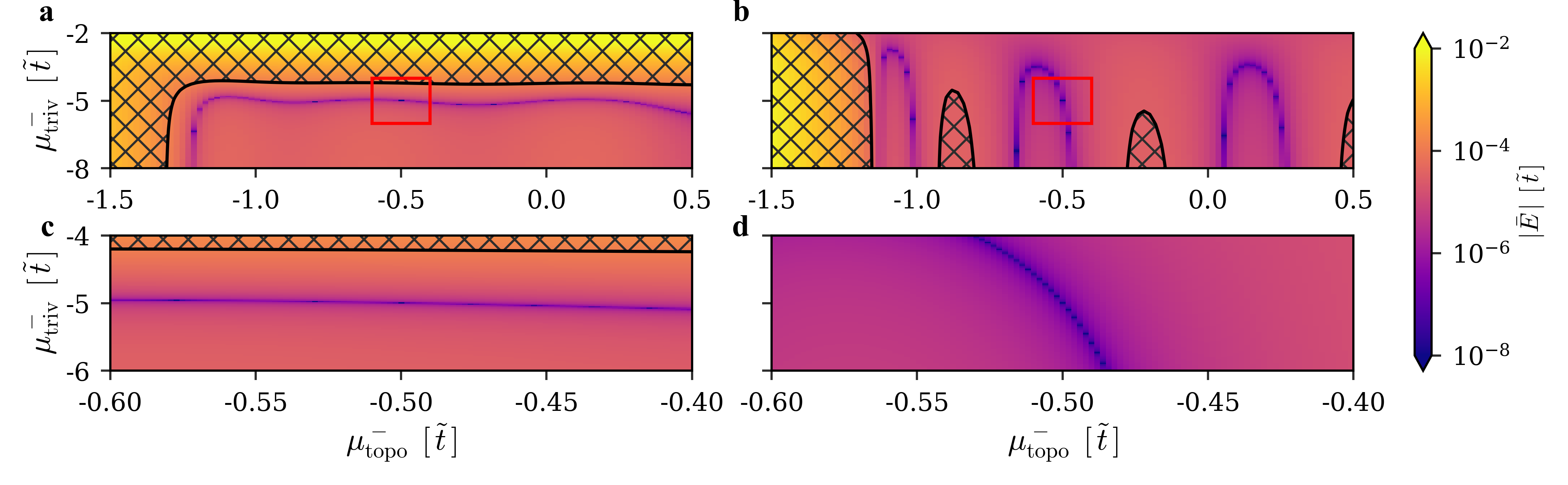}
        \caption{\textbf{Effects of imperfect calibration of negative chemical potentials on $\mathbf{\bar{E}}$. a.} The magnitude of the average hybridization energy, $\bar{E}$, throughout a corrected $\sqrt{X}$ gate for various values of the negative chemical potentials $\mu_{\rm topo/triv}^-$. The values of $\mu_{\rm topo}^-=-0.5\tilde{t}$ and $\mu_{\rm triv}^-=-5\tilde{t}$ (center of the red boxes) generate the perfect braid. The hatched regions denote where $\bar{E}$ for the original $\sqrt{X}$ is less than the corrected gate, when performed at the same speed. \textbf{b.} The same as \textbf{a} but for a $\sqrt{Z}$ gate. \textbf{c.} and \textbf{d.} Zoom into the small regions enclosed by the red boxes in \textbf{a} and \textbf{b}.}
        \label{fig:Ebar vs Neg mu}
    \end{figure}
    
To extend the analysis, Fig.\,\ref{fig:Ebar vs Neg mu}\,\textbf{a} and \textbf{b} show the magnitudes of $\bar{E}$ for the corrected $\sqrt{X}$ and $\sqrt{Z}$ gates, varying both $\mu^-_{\rm topo}$ and $\mu^-_{\rm triv}$. The points corresponding to the perfect swap ($\mu^-_{\rm topo} = -0.5\tilde{t}$, $\mu^-_{\rm triv} = -5\tilde{t}$) for each gate have $\bar{E}\approx0$ as expected, leading to the ideal results previously seen. Furthermore, instead of being isolated, these points lie on curves through the parameter space which also share a negligible value of $\bar{E}$, suggesting a range of possible other choices in $\mu^-_{\rm topo}$ and $\mu^-_{\rm triv}$ that will produce braids of comparable fidelity.

The precise shape of the $\bar{E}=0$ curve strongly depends on the values of $\mu^+_{\rm topo}$ and $\mu^+_{\rm triv}$, and as the hybridization throughout the braid depends non-linearly on the chemical potentials, is hard to predict analytically. In general, as the $\sqrt{X}$ gate hybridizes through the trivial region, its $\bar{E}$ is more dependent on $\mu^-_{\rm triv}$ while the opposite is true for the $\sqrt{Z}$ gate. 

The regions with black hatching shows where the plotted average hybridization exceeds that of the uncorrected gate. For all of the braids performed outside of those regions, it is expected that they will have a lower oscillation frequency, improving on the uncorrected gate. Within the red rectangles, centered on the ideal chemical potentials $\mu^-_{\rm topo} = -0.5\tilde{t}$, $\mu^-_{\rm triv} = -5\tilde{t}$, there is a significant portion of the space that is left un-hatched. Fig.\,\ref{fig:Ebar vs Neg mu}\,\textbf{c} and \textbf{d} show the respective regions with greater resolution.

\section{Effects of Random Disorder on Negative Hybridization in Symmetric Protocol}

Physical realizations of the relevant systems are unlikely to be perfectly clean; the presence of impurities and imperfections within the device can disrupt what should be perfectly uniform regions, potentially affecting the efficacy of negative hybridization in our proposed routine. To investigate these effects, we model disorder as a random perturbation to the chemical potential of each site, sampled from a uniform distribution on the interval $[-\frac{w}{2},\frac{w}{2})$ where we define $w$ as the disorder strength. These additions are time-independent, remaining constant through the entire duration of both the corrected and uncorrected braids.

    \begin{figure}[!h]
    \centering
    \includegraphics[width=0.98\linewidth]{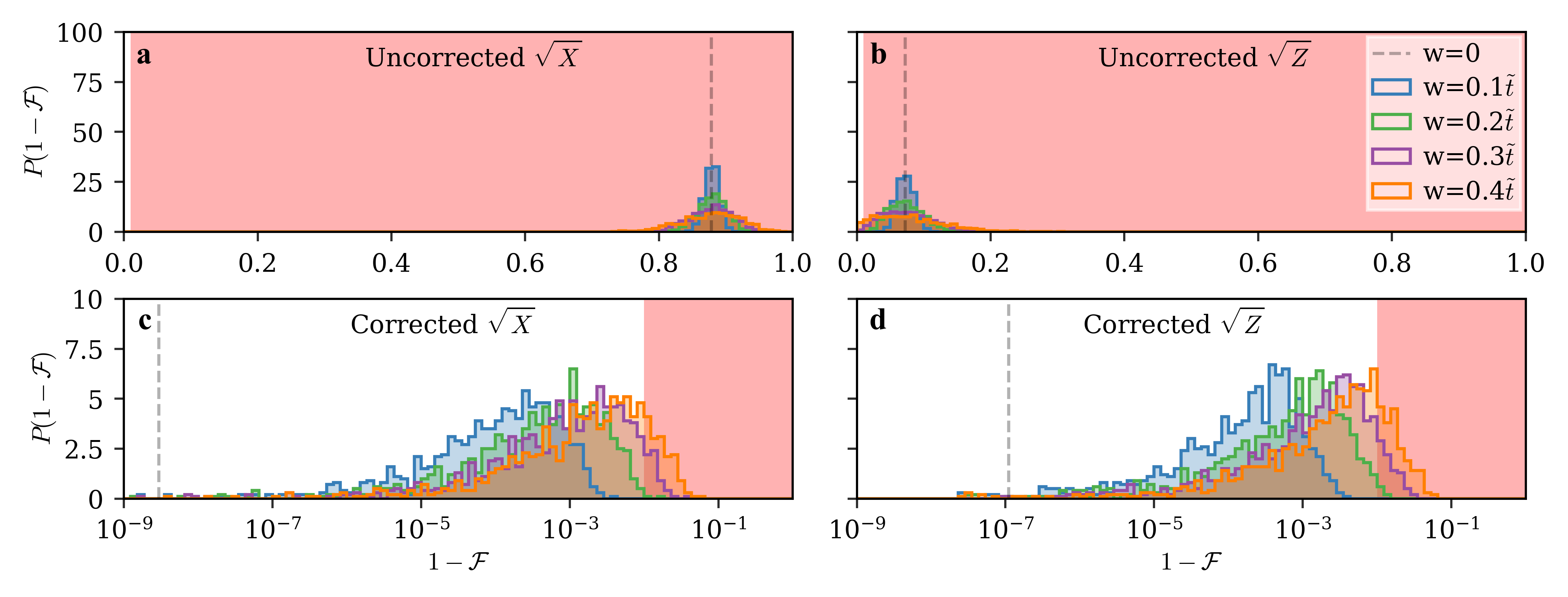}
    \caption{\textbf{Effect of disorder on symmetric braids a.} The probability distribution of 1000 $\sqrt{X}$ gates, each performed with a different random uniform disorder configuration, for various disorder strengths, $w$, as a function of the error, $1-\mathcal{F}$. The dashed line marks the error of the clean case where $w=0$, and the red shaded region is where the error exceeds the error correction threshold. \textbf{b-d.} The same as \textbf{a} for the $\sqrt{Z}$, corrected $\sqrt{X}$, and corrected $\sqrt{Z}$ gates, respectively.}
    \label{fig:Disorder}
    \end{figure}

Fig.\,\ref{fig:Disorder} shows the probability distributions of the error in corrected and uncorrected $\sqrt{X}$ and $\sqrt{Z}$ gates at braid time $T=20000\hbar/\tilde{t}$, measured as $1-\mathcal{F}$. The uncorrected gates in Fig.\,\ref{fig:Disorder}\,\textbf{a} and \textbf{b} are plotted on a linear scale, while the corrected gates in Fig.\,\ref{fig:Disorder}\,\textbf{a} and \textbf{b} are plotted on a logarithmic scale due to their smaller errors. The red shaded region in all plots indicates errors which exceed the error correction threshold.

1000 braids are simulated, each with a unique spatial disorder configuration, scaled to various values of $w$. The distributions for the uncorrected gates 
stay centered around the error of their clean cases, marked by the dashed lines, and spread out as $w$ increases. Meanwhile, the presence of disorder increases the mean error for the corrected gates by orders of magnitude, with stronger disorder strengths leading to larger errors on average.

These are due to the disorder altering how the MZMs hybridize, changing $\bar{E}$ for each basic braid. The random disorder breaks the spatial symmetry between the left and right legs of a T-junction, and causes the attempted transition to the negative chemical potential to not be perfectly accurate, both of which potentially reduce the effectiveness of the proposed correction to the standard braids.


Despite the fact that disorder disrupts the protocol, we can see that the distribution for the corrected gates remains significantly better than the uncorrected ones, even for large disorder strengths. Additionally, many different disorder configurations lead to corrected braids with fidelities still under the error correction threshold, and are therefore still usable for quantum computation. Specifically, all 1000 configurations with disorder strength $w=0.1\tilde{t}$ (and almost all configurations with $w=0.2\tilde{t}$) produced braids with error below below the fault tolerance threshold, see Fig.\,\ref{fig:Disorder}\,\textbf{c} and \textbf{d}.


\section{Parity swaps in Spinful Topological Superconductors}
\label{sec: spinful systems}

In the main paper, we focus on the Kitaev chain as the low energy limit for all MZM systems, however the structure in the parity of a system's ground state within parameter space is dependent on the specific Hamiltonian. Here, we briefly examine how the parity for spinful systems differ to that of the Kitaev chain, and look at the implications to the applicability of our proposed mechanism for correcting the hybridization within braids.

Common proposals for physical realizations of MZM hosting systems include a magnetic field to break the spin degeneracy, and either feature an intrinsic $p$-wave superconducting term\,\cite{setiawanTransportSuperconductorNormal2017}, or the more physical combination of Rashba spin-orbit coupling and $s$-wave superconductivity to induce effective $p$-wave superconductivity\,\cite{oregHelicalLiquidsMajorana2010a,lutchynMajoranaFermionsTopological2010}. The non-superconducting part of the spinful Hamiltonian for a chain with $N$ sites is given as follows,
\begin{equation}
    H_{0} = \sum^{N}_{j,s,s'} -\mu \sigma^0_{ss'} c_{j,s}^\dagger c_{j,s'}^{\phantom{\dagger}}+V_Z\sigma^z_{ss'} c_{j,s}^\dagger c_{j,s'}^{\phantom{\dagger}}+\sum_{j,s,s'}^{N-1}-\tilde{t} \sigma^0_{ss'} c_{j+1,s}^\dagger c_{j,s'}^{\phantom{\dagger}}
    -i\alpha_R\sigma^y_{ss'} c_{j+1,s}^\dagger c_{js'}^{\phantom{\dagger}}
    +{\rm H.c.}\,,
\end{equation}
where $V_Z$ is the Zeeman field strength, $\alpha_R$ is the Rashba spin-orbit coupling strength, $\sigma^\nu$ are the Pauli matrices acting on spin space, and as defined earlier, $\mu$ and $\tilde{t}$ are the chemical potential and hopping strength.

Adding the $p$-wave superconducting term, $\Delta_p$, produces the spinful Kitaev chain with intrinsic $p$-wave superconductivity,
\begin{equation}
    H_{p} = H_{0} + \sum_{j}^{N-1}\Delta_p\left( c_{j+1,\uparrow}^\dagger c_{j,\uparrow}^\dagger-c_{j+1,\downarrow}^\dagger c_{j,\downarrow}^\dagger\right)+{\rm H.c.}\,,
\end{equation}
while adding the $s$-wave superconducting term, $\Delta_s$, gives the spinful Kitaev chain with induced $p$-wave superconductivity,
\begin{equation}
    H_{s} = H_{0} + \sum_{j}^{N}\Delta_s \left(c_{j,\uparrow}^\dagger c_{j,\downarrow}^\dagger-c_{j,\downarrow}^\dagger c_{j,\uparrow}^\dagger\right)+{\rm H.c.}\,.
\end{equation}

\begin{figure}[!h]
    \centering
    \includegraphics[width=0.98\linewidth]{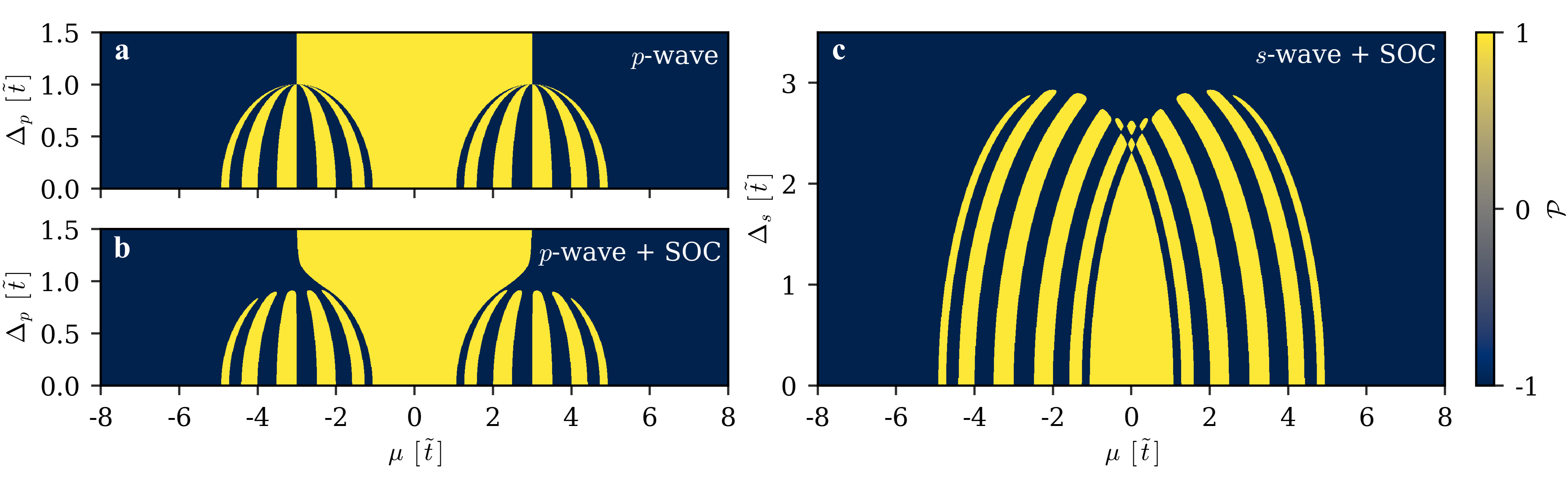}
    \caption{\textbf{Parity for spinful topological superconductor models. a.} The parity of the ground state of an eleven site spinful chain, as given by $H_p$, is plotted for different values of $\mu$ and $\Delta_p$. The other parameters are fixed at $V_z=3\tilde{t}$ and $\alpha_R=0$. The yellow regions show where $\mathcal{P}=1$ while the blue regions show where $\mathcal{P}=-1$. \textbf{b.} The same as \textbf{a}, but with finite Rashba spin-orbit coupling: $\alpha_R=0.1\tilde{t}$. \textbf{c.} The same as parameters as \textbf{b}, except now for the $H_s$ Hamiltonian, where the $p$-wave superconductivity is now induced by the $s$-wave order parameter, $\Delta_s$.}
    \label{fig:spinful_parity_dome}
\end{figure}

Fig.\,\ref{fig:spinful_parity_dome} shows the parity of the ground states (yellow for $\mathcal{P}=1$ and blue for $\mathcal{P}=-1$) across different values of chemical potential and superconducting gaps for a spinful chain with 11 sites and $V_Z=3\tilde{t}$. Fig.\,\ref{fig:spinful_parity_dome}\,\textbf{a} is the most direct comparison to the spinless Kitaev chain as it features the intrinsic $p$-wave superconductivity and $\alpha=0$. As such, the structure closely mimics the results in Fig.\,\ref{fig:MBS}\,\textbf{c}. The duplication of the dome-like feature means that the spinful system is now symmetric around $\mu=0$, in contrast to the original anti-symmetry. However, around the center of each dome ($\mu/\tilde{t}=\pm3$) the desired anti-symmetry is preserved. Thus by adapting the protocol to swap the chemical potentials around the center of each topological region we can regain the previous cancellation.

To determine how robust these structures are, we include a finite spin-orbit coupling of $\alpha=0.1\tilde{t}$ in Fig.\,\ref{fig:spinful_parity_dome}\,\textbf{b}. This causes some of the features to be warped, most notably around $\Delta_p/\tilde{t}=1$, breaking the desired anti-symmetry. However, for lower (and higher) superconducting strengths, the difference between \textbf{a} and \textbf{b} is small, suggesting that for many choices of parameters, swapping the chemical potential would still swap the parity and therefore lead to some reduction in $\bar{E}$, similar to the imperfect swaps discussed in Sec.\,\ref{Effects of introducing asymmetry}.

These warping effects are magnified for the chain with only intrinsic $s$-wave superconductivity, demonstrated in Fig.\,\ref{fig:spinful_parity_dome}\,\textbf{c} which still has $\alpha=0.1\tilde{t}$ and varies $\Delta_s$. For low values of $\Delta_s$ the anti-symmetry remains approximately preserved, but as it increases, the entire pattern quickly bends towards $\mu=0$.
 
For all of the examined systems, the parity of the ground state changes multiple times within the range of topological parameters, demonstrating that the presence of negative hybridization can be extended beyond the Kitaev chain limit.

\bibliography{Neghyb}